\documentclass[a4paper,10pt,aps,prx,twocolumn,superscriptaddress,floatfix]{revtex4-1}

\usepackage{lmodern}
\usepackage{float}
\usepackage{mathrsfs}
\usepackage{amsmath}
\usepackage{amsthm}
\usepackage{amssymb}
\usepackage{amsbsy}
\usepackage{graphicx}
\usepackage{bm}
\usepackage{hyperref}
\usepackage{cleveref}
\usepackage{lipsum}
\usepackage{sidecap}
\usepackage{epsfig}
\usepackage{array}
\usepackage[toc,page]{appendix}
\usepackage{tikz}
\usepackage{color}
\usetikzlibrary{intersections,calc}

\DeclareMathOperator{\sgn}{sgn}
\newcommand{\ket}[1]{\vert #1\rangle}
\newcommand{\bra}[1]{\langle #1\vert}

\begin{document}
\title{Giant Effective charges and Piezoelectricity in Gapped Graphene }
\author{Oliviero Bistoni}
\affiliation{Sorbonne Universit\'e, CNRS, Institut des Nanosciences de Paris, UMR7588, F-75252, Paris, France}
\affiliation{Dipartimento di Fisica, Universit\`a di Roma La Sapienza, Piazzale A. Moro 5, I-00185 Roma, Italy}
\affiliation{Graphene Labs, Fondazione Instituto Italiano di Tecnologia, Via Morego, I-16163 Genova, Italy}

\author{Paolo Barone}
\affiliation{Istituto Superconduttori, Materiali Innovativi e Dispositivi (SPIN-CNR), c/o Universit{\`a} G. D'Annunzio, I-66100 Chieti, Italy}

\author{Emmanuele Cappelluti}
\affiliation{Istituto di Struttura della Materia (ISM-CNR), I-34149 Trieste, Italy}

\author{Lara Benfatto}
\affiliation{Istituto dei Sistemi Complessi (ISC-CNR), UOS  Sapienza,  Piazzale  A. Moro  5,  I-00185  Roma,  Italy}
\affiliation{Dipartimento di Fisica, Universit\`a di Roma La Sapienza, Piazzale A. Moro 5, I-00185 Roma, Italy}

\author{Francesco Mauri}
\affiliation{Dipartimento di Fisica, Universit\`a di Roma La Sapienza, Piazzale A. Moro 5, I-00185 Roma, Italy}
\affiliation{Graphene Labs, Fondazione Instituto Italiano di Tecnologia, Via Morego, I-16163 Genova, Italy}
\begin{abstract}
Since the first realization of reversible charge doping in graphene via field-effect devices, it has become evident how the induction a gap could further enhance its potential for technological applications. Here we show that the gap opening due to a sublattice symmetry breaking has also a profound impact on the polar response of graphene. By combining ab-initio calculations and analytical modelling we show that for realistic band-gap values ($\Delta\lesssim 0.5$ eV) the piezoelectric coefficient and the Born effective charge of graphene attain a giant value, independent on the gap. In particular the piezoelectric coefficient per layer of gapped mono- and bilayer graphene is three times larger than that of a large-gap full polar insulator as hexagonal Boron Nitride (h-BN) monolayer, and 30\% larger than that of a polar semiconductor as MoS$_2$. This surprising result indicates that piezoelectric acoustic-phonons scattering can be relevant to model charge transport and charge-carrier relaxation in gated bilayer graphene. The independence of the piezoelectric coefficient and of the Born effective charge on the gap value  follows from the connection between the polar response and the valley Chern number of gapped Dirac electrons, made possible by the effective gauge-field description of the electron-lattice/strain coupling in these systems. In the small gap limit, where the adiabatic ab-initio approximation fails, we implement analytically the calculation of the dynamical effective charge, and we establish a universal relation between the complex effective charge and the so-called Fano profile of the phonon optical peak. Our results provide a general theoretical framework to understand and compute the polar response in narrow-gap semiconductors, but may also be relevant for the contribution of piezoelectric scattering to the transport properties in Dirac-like systems. 
\end{abstract}

\maketitle

\section{Introduction}
After more than ten years from the discovery of exfoliated monolayers\cite{Novoselov2004}, graphene is still object of intense research activity due to its fascinating electrical, mechanical, and optical properties. The unique sublattice symmetry arising from its 2D honeycomb lattice leads to a gapless Dirac cone structure at the corners of the corresponding Brillouin zone that is responsible for its peculiar electronic properties. Pristine graphene displays therefore a semimetallic behavior and, as such, it cannot sustain any static polarization. Nonetheless, several symmetry-breaking schemes have been proposed and adopted to engineer a moderate bandgap in graphene, among which we cite disorder~\cite{D'Apuzzo2017}, confined geometries such as quantum dots or nanoribbons~\cite{Geim2007,GNR.PhysRevLett.97.216803,CHEN2007228,PhysRevLett.98.206805}, interaction with the substrate~\cite{Zhou2007,Novoselov2007,Nevius_prl2015} and applied electric field in bilayer graphene~\cite{NMat7.2007, Kuzmenko_prl09,Feng_naturenano09}. 

When a bandgap is open, and once the proper symmetry requirements are met, internal polarizations are in principle allowed in gapped graphene, leading  to piezoelectricity or to infrared (IR) absorption of its optically-active phonon modes, as well as mediating piezoelectric and polar (Fr\"ohlich) electron-phonon interactions with foreseeable consequences on transport properties and on relaxation dynamics of photo-carriers. So far, there have been proposals to engineer piezoelectricity in graphene by, e.g., creating triangular holes in dielectric graphene nanoribbons\cite{SharmaAPL2012} or by chemical doping of graphene sheets~\cite{Ong2012,Ong2013}. Even though both schemes entail a breaking of the inversion symmetry and the opening of a bandgap, the predicted piezoelectric constants were found to be one order of magnitude smaller than those recently computed for other non-centrosymmetric 2D hexagonal crystals, such as hexagonal Boron Nitride (h-BN) and transition-metal dichalcogenides (TMDs)\cite{DuerlooJPCL2012,Droth_PRB2016, Rostami_njp2d2018,piezo2d_review}. On the other hand, a giant increase of the IR intensity has been explicitly observed in gated bilayer graphene as a function either of the gate voltage\cite{Kuzmenko_prl09} or of the gap size\cite{Feng_naturenano09}. The resulting phonon peak in the optical conductivity displays a pronounced Fano-like asymmetry, and both phenomena have been ascribed to a ``charged phonon'' effect, where the IR activity is borrowed by the electronic background interacting with the phonon mode\cite{Cappelluti_prb10,Cappelluti_prb12}. Interestingly, also optical measurements on nanoporous graphene revealed a clear spectroscopic feature close to the frequency of the graphene $E_{2g}$ phonon mode, whose IR activity has been suggested to arise from disorder of the nanoporous structure\cite{D'Apuzzo2017}.

In this work, we systematically study the possibility of engineering sizeable internal polarization in gapped graphene by considering different and experimentally achievable symmetry-breaking perturbations. To this end, we calculate from first principles -- in the framework of Density Functional Theory (DFT) -- two physical quantities which describe the coupling between the electric field and the vibrational modes of an insulating material, namely the piezoelectric tensor and the Born effective charges in the static limit. The piezoelectric tensor represents the coupling of the electric field to the cell deformations, or, stated differently, with the acoustic phonon modes when $\textbf{q}\rightarrow 0$, thus being related to the piezoelectric electron-phonon interaction of acoustic phonons.
On the other hand, Born effective charges allow one to quantify the coupling between the electric field and the optical phonon modes in the long-wavelength limit ($\textbf{q}\rightarrow 0$), i.e., to assess the IR vibrational spectra as well as the Fr\"ohlich electron-phonon interaction. 

Interestingly, our calculations unveil a universal behavior of both piezoelectric and Born effective charge tensors in gapped graphene and gated bilayer graphene. When the external perturbation breaks the equivalence between the two carbon atoms of the unit cell (or, stated differently, the sublattice symmetry), gapped graphene monolayer displays Born effective charges and piezoelectric constants comparable to those of gated bilayer graphene. In addition, in both systems these two quantities are substantially independent both on the bandgap width $\Delta$ (at least in the experimentally accessible range $\Delta \lesssim 0.5~eV$) and on the Fermi velocity. The predicted Born effective charges are comparable with those of h-BN monolayer, a strongly polar material, while the piezoelectric constant is 3 times larger than that computed ab-initio for h-BN\cite{DuerlooJPCL2012}, and 30$\%$ larger than that of a polar semiconductor as MoS$_2$. This surprising result indicates that piezoelectric acoustic-phonon scattering can be relevant to model charge transport and charge-carrier relaxation in gated bilayer graphene.
 To get an analytical insight on the origin of these giant universal values we  derive  both quantities in gapped graphene monolayer using the modern theory of polarization\cite{PhysRevB.48.4442,RevModPhys.66.899} and a modified Dirac-like model which includes the sublattice-symmetry breaking. Within this framework the coupling between the electrons and the lattice/atomic distortions can be expressed through  appropriate gauge fields\cite{Rostami_njp2d2018,PhysRevB.90.125414,110005716689,200662175,doi:10.1143/JPSJ.74.777,PhysRevB.65.235412,Piscanec04,PhysRevB.79.165431}. Then we show that not only the piezoelectric constant\cite{Rostami_njp2d2018} but also the in-plane Born effective charge can be expressed in terms of the valley Chern number, which is independent on the gap value and is proportional solely to the gauge-field coupling constants. This result allows one to explain why gapped graphene displays the same giant polarization as an inherently polar material like h-BN monolayer. 
As the gap value decreases the  static (adiabatic) approximation is expected to fail. Regarding the Born effective charges this occurs when the  bandgap is smaller than the optical phonon energy, requiring a complete frequency-dependent calculation. Here we provide an analytical estimate of the dynamical effective charges  within the gapped Dirac-like model. Our derivation, that is shown to be fully consistent with the diagrammatic field-theory approach adopted for gated bilayer graphene\cite{Cappelluti_prb10,Cappelluti_prb12}, demonstrates that the IR activity not only vanishes ad expected as $\Delta \rightarrow 0$, but it also displays a huge increase at resonance, i.e. when the bandgap is comparable with the frequency of the optically-active phonon. Furthermore, we relate the Fano asymmetry of the phonon peak to the imaginary part of the (complex) dynamical effective charge, in analogy with the findings of Refs. \onlinecite{Cappelluti_prb10,Cappelluti_prb12}.

The paper is organized as follows. Sec.~\ref{section:theory} provides the theoretical background, defining the basic quantities that will be analyzed in the following sections and briefly discussing their limits of validity and applicability. In Sec. ~\ref{Sec:sym_break} we review the symmetry requirements that both the piezoelectric and the Born effective charge tensors must obey and we introduce the symmetry-breaking mechanisms that will be considered, while in Sec. \ref{Sec:methods} we illustrate the computational and analytical models adopted to describe them. Piezoelectricity and (static) Born effective charges in gapped graphene are discussed in Secs. \ref{Sec:piezo},\ref{Sec:BEC}, while Secs. \ref{Sec:model}, \ref{Sec:optical} are devoted to the frequency-dependent calculation of effective charge within the Dirac-cone approximation, and its impact on the optical conductivity of gapped graphene.

\section{Theoretical background}\label{section:theory}

The Born effective charge  tensor $Z_{\alpha\beta}^I$ and the piezoelectric tensor $e_{\alpha\beta\gamma}$ are defined as
\begin{equation}
 Z_{\alpha\beta}^I=\frac{\Omega}{|e|}\frac{\partial P_\alpha}{\partial u_\beta^I},	\quad e_{\alpha\beta\gamma}=\frac{\partial P_\alpha}{\partial\epsilon_{\beta\gamma}},	\label{Eq:definitions}
\end{equation}
where $e$ is the electronic charge, $u_\beta^I$ is the static atomic displacement of ion $I$ along the $\beta$ direction and $\epsilon_{\beta\gamma}$ is the strain tensor, the Greek indexes used hereafter for Cartesian components.
In 3D systems $P_\alpha$ is the macroscopic polarization and $\Omega$ is the unit cell volume, while for 2D systems $P_\alpha$ is the dipole moment per unit surface and $\Omega$ is the unit cell area.

Since the definitions of Eq.~(\ref{Eq:definitions}) are static, they cannot be applied to metallic systems, for which a static polarization cannot be defined. While in metals the effective charges and piezoelectric tensor can be defined only with perturbations of finite frequency, the static definitions given in Eq.~(\ref{Eq:definitions}) can be safely adopted to characterize the lattice response of insulators to an applied electric field as long as the vibrational frequencies are much smaller than the electronic bandgap. The piezoelectric tensor is probed by quasi-static strain deformations or, stated differently, by acoustic low-energy phonons. Therefore, a static approximation is fully justified even in narrow-gap semiconductors. 
{\color{black}On the other hand, the latter may display electronic excitation energies that are comparable with the optical phonon frequencies. This is particularly true in (gapped) graphene systems, that exhibit phonon energies up to $0.2~eV$ and bandgap of comparable or smaller size. In order to evaluate the optical response under such circumstances, it is therefore necessary to generalize the static definition of the Born effective charge} by considering the derivative of the polarization with respect to a time-dependent phonon displacement oscillating at frequency $\omega$.
Within a DFT scheme, the dynamical effective charge tensor may be decomposed into ionic and electronic contributions:
\begin{equation}
 Z_{\alpha\beta}^I(\omega)=Z^I\delta_{\alpha\beta}+Z^{I,el}_{\alpha\beta}(\omega).
 \label{Eq:Ztot}
 \end{equation}
The ionic contribution $Z^I$ is equal to the ionic charge {\color{black}in a pseudopotential calculation, where the core electrons are assumed to follow rigidly the nuclei displacement. }   The  $Z^{I,el}$ accounts for polarization of the valence electrons induced by the lattice vibrations, and it can be evaluated by means of the time-dependent Density Functional Perturbation Theory\cite{GROSS1990255} (DFPT) as:
\begin{multline}
 Z^{I,el}_{\alpha\beta}(\omega) = \frac{2}{N_k} \sum_{\mathbf{k}} \sum_{i,j} \ \frac{ \theta\left(\epsilon_F-\epsilon_{\mathbf{k}i}\right) - \theta\left(\epsilon_F-\epsilon_{\mathbf{k}j}\right) }{\epsilon_{\mathbf{k}i}-\epsilon_{\mathbf{k}j}+\hbar\omega+i\eta} \times \\
\times \left\langle u_{\mathbf{k}i}\right| \frac{ie\hbar v_{\alpha}}{\epsilon_{\mathbf{k}i}-\epsilon_{\mathbf{k}j}} \left| u_{\mathbf{k}j}\right\rangle  \left\langle u_{\mathbf{k}j}\right| \frac{\partial V^{KS}}{\partial u^I_\beta} \left| u_{\mathbf{k}i}\right\rangle,		\label{Eq:dynamic_charge}
\end{multline}
where $|u_{\mathbf{k}i}\rangle$ is the periodic part of the unperturbed Bloch wave function with Kohn-Sham energy $\epsilon_{\mathbf{k}i}$ and band index $i$,
 $V^{KS}$ is the Kohn-Sham potential, $\epsilon_F$ is the Fermi energy, $\theta$ is the Heaviside step-function, $\eta$ is a small
positive real number, and  $v_\alpha$ is the velocity operator, defined as 
\begin{equation}
v_\alpha =\frac{1}{\hbar}\frac{\partial H_\textbf{k}^{KS}}{\partial k_\alpha}.
\label{Eq:velocity}
\end{equation}
In semiconductors, the frequency-dependent Born effective charge Eq.~(\ref{Eq:dynamic_charge})  is a complex quantity if $\hbar\omega$ is larger than the direct gap.  As we will discuss in Sec. \ref{Sec:optical}, the complex nature of the frequency-dependent effective charge will largely impact the optical response of narrow-gap semiconductors. On the other hand, if $\hbar\omega$ is smaller than the direct gap, $Z^{I,el}_{\alpha\beta}(\omega)$ is a real quantity and reduces to the static limit for $\omega=0, \eta=0$, that is routinely used to assess the optical strength of the IR vibrational response\cite{Giannozzi_IR1994}.

\section{Symmetry breaking}	\label{Sec:sym_break}

Before computing explicitly the Born effective charge and piezoelectric tensors, it can be useful to review their symmetry requirements in graphene and related systems, and to  describe the symmetry-breaking schemes that will be analyzed and discussed in the next sections.

\begin{figure*}
\centering
\includegraphics[scale=1.25]{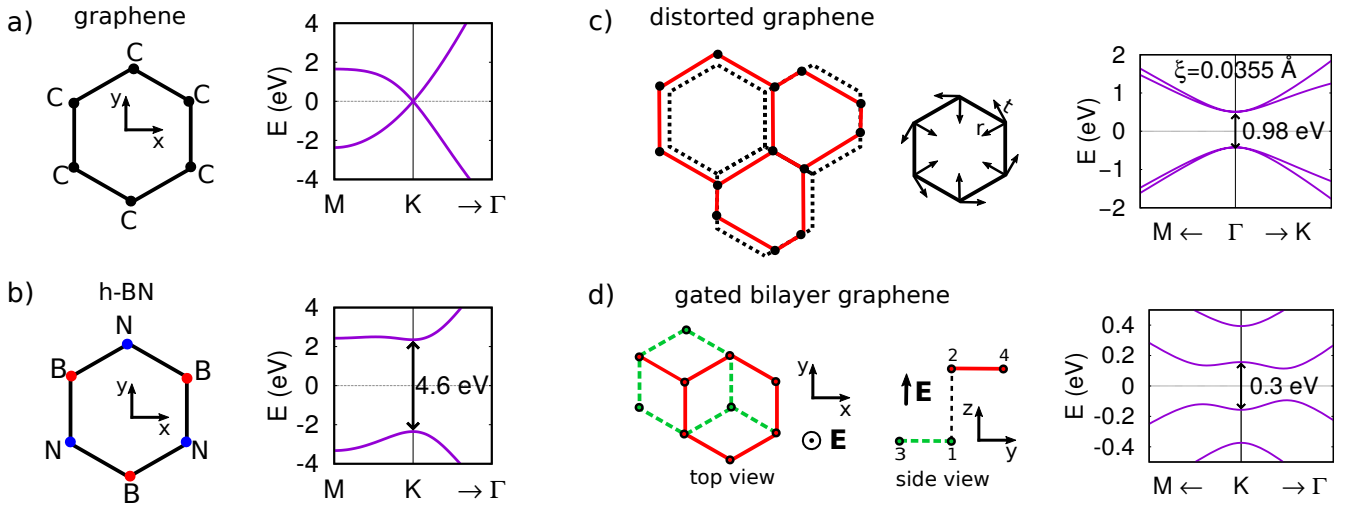}
\caption{ Atomic structure and band structure of graphene, hexagonal Boron Nitride (equivalent to disproportionated graphene from a symmetry point of view), distorted graphene and gated bilayer graphene. a) Graphene is a semimetal with two Carbons per cell and a linear dispersion at point $K$. b) Monolayer hexagonal Boron Nitride is a wide gap insulator isomorphic to graphene with one atom of Boron (red) and one of Nitrogen (blue) per cell; its diatomic nature, breaking the sublattice symmetry characteristic of graphene, causes a gap opening at valley $K$ (the estimated bandgap at DFT level is also given). c) Distorted graphene (solid red line) is obtained, with respect to pristine graphene (dashed black line), as a shifting of the atoms in the radial direction every third hexagon. The sign of the shifting $\xi$ is assumed to be positive for the expansion and negative for the shrinking of the hexagon. Radial and tangential directions are identified by the $r$ and $t$ axis. The band structure exhibits a gap at $\Gamma$ whose amplitude (and sign) depends on the strength (and sign) of the deformation: we get $\Delta\simeq 0.98$~eV for $\xi=0.0355$~\AA. d) The structure of bilayer graphene is shown from top and side view. Green dashed and red solid lines represent the lower and the upper plane, respectively. The electric field is perpendicular to the atomic planes. The amplitude of the gap is related to the intensity of the electric field. We assume that the gap is positive for $\mathbf{E}=|E|\hat z$ and negative for $\mathbf{E}=-|E|\hat z$. The gap width is computed as the difference between conduction and valence bands at K point.}	\label{Fig:symmetrybreaking}
\end{figure*}

Graphene is a semimetal with 2 Carbons per unit cell, arranged in a 2D hexagonal lattice with layer group $p6/mmm$, whose electronic band structure exhibits linear dispersion close to the $K$ ($K'$) points of the Brillouin zone (see Fig. \ref{Fig:symmetrybreaking}a). Since the crystallographic point group $D_{6h}$ admits a center of inversion, the piezoelectric tensor is identically zero. On the other hand, the effective charge tensor is, in principle, allowed to be nonzero, the lattice symmetries imposing it to be diagonal with two independent in-plane and out-of-plane components $Z^I_{xx}=Z^I_{yy}\equiv Z^I_\parallel$ and $Z^I_{zz}=Z^I_\perp$ for each carbon atom. Since the two carbons in the unit cell are equivalent, their effective charges are the same; nonetheless, the acoustic sum rule (ASR) stemming from translational invariance and charge neutrality\cite{ASR_PRB1970} requires that $\sum_I Z^{I}_{\alpha\mu}=0$, implying that the effective charge tensor must vanish. 
This is consistent with the fact that no IR activity is allowed in pristine graphene, the only optically active phonon mode being the Raman-active, in-plane $E_{2g}$ mode at 1582~$\text{cm}^{-1}$.

The crystal symmetry is reduced to $p\bar{3}m1$ in bilayer $AB$ Bernal-stacked graphene\cite{PhysRevB.79.125426},
where half of the carbon atoms lie over an atom in the neighboring graphene sheet, the other half lying directly over the center of a hexagon in the next layer, as shown in Fig. \ref{Fig:symmetrybreaking}d. The interlayer hybridization modifies the linear dispersion at the $K$ point, turning bilayer graphene in a zero-gap semiconductor with parabolic valence and conduction bands touching at valleys $K$ and $K'$. The crystallographic point group $D_{3d}$ of bilayer graphene still admits a center of inversion and, hence, piezoelectricity is not allowed. The Born effective charges of bilayer graphene, instead, are nonzero because the carbon atoms belonging to the two sublattices of each layer, e.g., atoms 1 and 3 shown in the side view scheme of Fig.~\ref{Fig:symmetrybreaking}d, are not equivalent.  As for graphene monolayer, the lattice symmetry imposes the effective charge tensor to be diagonal, with two independent $Z_\parallel$ and $Z_\perp$ components for each carbon atom. Inversion symmetry relates atoms 1 (3) and 2 (4), located on different graphene sheets, which therefore display the same effective charges. Finally, imposing the charge-neutrality ASR one obtains $Z^{C1}_{\parallel(\perp)}=Z^{C2}_{\parallel(\perp)}=-Z^{C3}_{\parallel(\perp)}=-Z^{C4}_{\parallel(\perp)}$. 
As for the vibrational optical activity, and consistently with the non-zero Born effective charges, the crystal symmetry allows for two IR-active optical modes, stemming from out-of-phase lattice displacements in the two graphene sheets, namely an in-plane two-fold degenerate $E_{u}$ mode and an out-of-plane $A_{2u}$ mode, analogously to graphite\cite{graphite_IR}.

It is useful to consider also the simplest example of a diatomic crystal with 2D hexagonal lattice, namely h-BN monolayer. h-BN monolayer is isomorphic to graphene but, displaying two inequivalent atoms per cell, it belongs to a different layer group $p\bar{6}2m$ (crystallographic point group $D_{3h}$), hence lacking inversion symmetry. Due to its diatomic nature, h-BN monolayer is insulating with a large direct gap of approximatively 6$~eV$ at $K$ point\cite{Watanabe2004}.
Having no centre of inversion, it may exhibit piezoelectricity; theoretical predictions suggest, in fact, that piezoelectric constants of h-BN monolayer should be quite large and comparable to those of popular bulk piezoelectric\cite{DuerlooJPCL2012, Droth_PRB2016, Rostami_njp2d2018}.
Point group symmetry $D_{3h}$ implies a single independent piezoelectric coefficient, being
$e_{222}=-e_{211}=-e_{121}=-e_{112}$,
where 1,2 represent $x$ and $y$ of the chosen Cartesian reference frame shown in Fig. \ref{Fig:symmetrybreaking}.
As for graphene, the Born effective charge tensor is diagonal, with two independent components $Z^I_\parallel$ and $Z^I_\perp$. Additionally, the ASR constrains the in-plane  and out-of-plane  Born effective charges to be opposite for B and N, i.e., $Z_{\parallel(\perp)}^B=-Z_{\parallel(\perp)}^N\equiv Z_{\parallel(\perp)}$. Consistently, two IR-active modes are allowed in h-BN, namely an in-plane two-fold degenerate $E^\prime$ mode and an out-of-plane $A_{2}^{\prime\prime}$ mode. 
Both the piezoelectric coefficient and Born effective charges have been calculated in the framework of DFT, with the reported values being $e_{222}=1.38\times10^{-10}~C/m$\cite{DuerlooJPCL2012}, $Z_\parallel = 2.7 $ and $Z_\perp = 0.8$\cite{PhysRevB.63.115207}.

In this paper we consider three different mechanisms to induce a symmetry breaking in graphene: {\emph a)} 
a sublattice symmetry breaking, i.e., a breaking of the equivalence between the two Carbon atoms (we refer to this case as disproportionated graphene); {\emph b)} a distortion of the hexagonal lattice inducing a $\sqrt{3}\times\sqrt{3}\text{R30}^\circ$ superstructure,
lowering the translational invariance and inducing inequivalent hopping channels between Carbon atoms; {\emph c)} an applied electric field on bilayer graphene in the out-of-plane direction. All these symmetry-breaking mechanisms induce a bandgap opening and a nonzero coupling between optical  modes and radiation, which is quantified by nonzero Born effective charges, whereas piezoelectricity is allowed only when the inversion symmetry is lost, as in the case of  {\emph a)} and {\emph c)} mechanisms. The modified symmetry of gapped graphene imposes new constraints on both Born effective charge and piezoelectric tensors, as illustrated below.

\paragraph{Disproportionated graphene.} The sublattice symmetry of graphene can be broken by the interaction with properly chosen substrates as, e.g., SiC\cite{Novoselov2007,Zhou2007,Nevius_prl2015}, where every second carbon atom has a neighbor in the bottom layer, a modulated potential is induced in graphene, that acts differently on atoms belonging to different sublattices. The equivalence between Carbon atoms is thus broken, with a reported band splitting up to 0.5 eV between the valence and conduction bands at the $K$ point\cite{Nevius_prl2015}. From a symmetry point of view, disproportionated graphene is equivalent to monolayer h-BN, and the piezoelectric and Born effective charge tensors obey exactly the same constraints. The piezoelectric tensor has a single independent coefficient $e_{222}$, while the Born effective charge tensor is diagonal with two independent components $Z_{\parallel(\perp)}$ describing in-plane (out-of-plane) Born effective charges. As for h-BN, the in-plane $E^\prime$ and out-of-plane $A_{2}^{\prime\prime}$ modes are IR-active.

\paragraph{Distorted graphene.} The hexagonal-lattice distortion with the $\sqrt{3}\times\sqrt{3}\text{R30}^\circ$ reconstruction,
where every third carbon hexagon in the graphene lattice expands or shrinks, lowers the translational symmetries of pristine graphene, as shown in Fig.~\ref{Fig:symmetrybreaking}c, while keeping the crystallographic point-group symmetry  . The crystal periodicity is modified accordingly, and the $\sqrt{3}\times\sqrt{3}\text{R30}^\circ$ superstructure is described by a unit cell with 6 carbon atoms rotated by 30$^\circ$ and rescaled by a factor $\sqrt{3}$ with respect to the primitive unit cell with 2 carbon atoms.
Signatures of this kind of distortion measured by scanning tunnelling microscopy and Raman spectroscopy have been reported for graphene nanoribbons and on the edges of graphite sheets (see, e.g., Ref.~\onlinecite{wassmann2010} and references therein).
In the supercell setting, the Dirac points of pristine graphene are folded to the $\Gamma$ point, where the distortion induces a bandgap opening, the bandgap width being related to the amount of the distortion. Since the point-group symmetries of distorted graphene are unchanged, the presence of a center of inversion implies that the piezoelectric tensor is identically zero.
Nonetheless, because of the translation-symmetry lowering, the in-plane Born effective charges of distorted graphene can be non-zero, being generally different from the ones of disproportionated graphene. Instead of the cartesian reference system, it is more convenient to adopt the radial-tangential coordinate system shown in Fig.~\ref{Fig:symmetrybreaking}c, where the Born effective charge tensor is diagonal. {\color{black}In this reference system, } the charge-neutrality ASR imposes for the in-plane components that $Z_r=-Z_t$, while the out-of-plane component is zero.
The distorted structure has its own vibrational modes: the only optical IR-active mode is the in-plane $E_{1u}$, corresponding to the $K_5$ eigenmode of pristine graphene\cite{PhysRevB.76.035439},  experimentally visible at 1218~$\text{cm}^{-1}$, folded to $\Gamma$ in the supercell setting.
\paragraph{Gated bilayer graphene.}
It is well known that gated bilayer graphene exhibits a gap at valleys $K$ and $K'$ whose width is related to the intensity of the electric field. The largest reported gate-induced bandgap that has been realized amounts to approximatively 250 meV~\cite{Zhang2009}. {\color{black} When the electric field is switched on, the equivalence between layers is broken because of the potential gradient along the $z$ direction, and inversion symmetry is lost.} As a consequence, the layer group reduces to $p3m1$ for perfectly perpendicular electric field, and the crystallographic point group reduces from $D_{3d}$ to $C_{3v}$; since $C_3$ and $\sigma_v$ symmetries are preserved, the piezoelectric tensor obeys similar constraints as in h-BN, being characterized by a single independent coefficient $e_{222}$. Similarly, the Born effective charge tensor remains diagonal, with two independent components $Z^I_\parallel$ and $Z^I_\perp$ per atom. However, the equivalence between carbon atoms 1 (3) and 2 (4), enforced by inversion symmetry in unbiased bilayer graphene, is lost (see Fig. \ref{Fig:symmetrybreaking}d), and the Born effective charges cannot be simply described by two global coefficients $Z_{\parallel(\perp)}$.
As for the vibrational optical activity, the inversion-symmetry breaking causes all optical phonon modes $3A_1+3E$ to be simultaneously Raman- and IR-active.

\section{Methods}\label{Sec:methods}

For each symmetry-breaking mechanism discussed in the previous section, we computed the piezoelectric coefficients and the Born effective charges  within the local-density approximation (LDA)\cite{PhysRevLett.45.566} of density functional theory (DFT) as implemented in Quantum ESPRESSO~\cite{0953-8984-21-39-395502}.  The relaxed-ion piezoelectric constant was estimated as $e_{222}=[P_2(\epsilon_{22})-P_2(-\epsilon_{22})]/2\vert\epsilon_{22}\vert$,
where $\epsilon_{22}$ is the uniaxial strain, ions were allowed to relax for each strained cell and $P_2$ is the surface polarization obtained from Berry phase calculation~\cite{PhysRevB.48.4442,RevModPhys.66.899}. The static effective charges have been computed {\sl ab-initio} by means of  density-functional perturbation theory (DFPT) \cite{RevModPhys.73.515}. Details of the calculations are given in App.~\ref{App:DFT}, while we discuss here the computational models we adopted to simulate the different symmetry-breaking mechanisms. 

{\em a. Disproportionated graphene.}
Graphene with inequivalent C ions can be seen as a continuous transformation from pristine graphene to h-BN monolayer. Such transformation can be modelled by varying the atomic charges of the basis atoms from $Z_{C1}=Z_{C2}=6$ (Carbons) to $Z_{C1}=6-\delta$ (Boron-like) and $Z_{C2}=6+\delta$ (Nitrogen-like) with $\delta$ between 0 and 1 and $2\delta$ measuring the charge disproportion between Carbon ions.
For every selected value of $\delta$, two pseudopotentials (one per atom) were generated with the package \textit{atomic} of Quantum ESPRESSO. All calculations were done, then, with the atomic masses and lattice parameter of pristine graphene ($a_0=2.460$~\AA), which is slightly smaller than the lattice parameter of h-BN monolayer ($2.504$~\AA). As expected, the sublattice-symmetry breaking opens a bandgap at $K$ point, whose width is found to increase almost linearly with $\delta$ from 0 to $\sim 4.6~eV$, which corresponds to the h-BN bandgap calculated within LDA.

{\em b. Distorted graphene.}
The lattice distortion has been simulated using a $\sqrt{3}\times\sqrt{3}\ \text{R30}^\circ$ supercell with 6 Carbons and modifying the atomic coordinates in the radial direction. 
We considered atomic displacements $\xi$ from the equilibrium positions of pristine graphene (dotted lines in Fig. \ref{Fig:symmetrybreaking}c), with step $\xi_0=8.875\times10^{-3}$~\AA, up to a maximum distortion corresponding to a bandgap of  $2~eV$.

{\em c. Gated bilayer graphene.}
The applied electric field was simulated by adding a saw-tooth potential along the $\hat{z}$ direction to the bare ionic potential. The slope of the potential determines the intensity of the electric field and thus the amplitude of the gap.
The chosen range of the slope variation allows for a bandgap tuning in a range between 0 and 0.4~$eV$, which is the physically significant range for gated bilayer graphene.
The corresponding electric field can then be obtained from the calculations of Ref.\onlinecite{PhysRevB.79.165431}.

In order to get further insight into our computational results, we also complemented our DFT study with an analysis of a modified Dirac-like model for disproportionated graphene. We start from the usual Dirac-like model $\mathcal{H}_{ K}({\bm k})=\hbar v_F{\bm k}\cdot\bm\sigma$, where $\hbar$ is the reduced Planck constant, $v_F$ is the Fermi velocity, ${\bm k}=(k_x,k_y)$ is the electron-momentum measured with respect to the Dirac point ${K}$, in a Cartesian basis, and $\bm\sigma=(\sigma_x,\sigma_y)$ is the Pauli matrix accounting for the sublattice isospin.
The inequivalence between Carbon ions can be modelled by a sublattice staggered potential which is diagonal in the basis of atomic sites $|A\rangle$ and $|B\rangle$ -- i.e., in the sublattice isospin -- and opens a gap $\Delta$ between the $\pi^\ast$ and $\pi$ bands, thus obtaining
\begin{equation}
 \mathcal{H}_\mathbf{K}(\mathbf{k})= \begin{pmatrix} \Delta/2& \hbar v_F(k_x-ik_y)\\ \hbar v_F(k_x+ik_y)& -\Delta/2 \end{pmatrix}.	\label{Eq:Hamiltonian}
\end{equation}
By diagonalizing the Hamiltonian (\ref{Eq:Hamiltonian}) one obtains the following eigenvalues and eigenfunctions for $\pi$ and $\pi^\ast$ bands:
\begin{subequations}
\label{Eq:eigenthings}
\begin{align}
 E_\mathbf{k}^{\pi^\ast}&=+E, \quad u_\mathbf{k}^{\pi^\ast}=\binom{u}{v},\\
 E_\mathbf{k}^\pi&=-E, \quad u_\mathbf{k}^\pi=\binom{-v^\ast}{u},
\end{align}
\end{subequations}
where $E=\sqrt{\Delta^2/4+(\hbar v_Fk)^2}$, $u$ and $v$ are given by
\begin{equation}
 u=\frac{1}{\sqrt{2}}\left(1+\frac{\Delta/2}{E}\right)^{\frac{1}{2}} \quad
 v=\frac{e^{i\theta}}{\sqrt{2}}\left(1-\frac{\Delta/2}{E}\right)^{\frac{1}{2}},	\label{Eq:coefficients}
\end{equation}
and $\theta=\tan^{-1}(k_y/k_x)$. 

The coupling of electrons with strain and lattice distortions can be accounted for by two different gauge fields, $\pmb{\mathcal{ A}}^{e-s}$ and $\pmb{\mathcal{ A}}^{e-l}$, respectively, defined as\cite{Piscanec04,Pisana2007,PhysRevB.90.125414, Droth_PRB2016, Rostami_njp2d2018, 110005716689,200662175,doi:10.1143/JPSJ.74.777,PhysRevB.65.235412}:  
\begin{eqnarray}
\pmb{\mathcal{A}}^{e-s} &=& (\epsilon_{11}-\epsilon_{22})\,\hat{\bm x}\,-2\,\epsilon_{12}\,\hat{\bm y} \label{eq:strain_field}\\
\pmb{\mathcal{A}}^{e-l} &=& \hat{\bm z}\times\bm u\equiv -u_y\,\hat{\bm x}\,+ u_x\,\hat{\bm y}\label{eq:lattice_field}
\end{eqnarray}
where $\epsilon_{\beta\gamma}$ is the strain tensor and $\bm u$ is the atomic displacement. 
The Hamiltonians describing the coupling with the two structural distortions can be written in a compact form as:
\begin{eqnarray}
\mathcal{H}_{K}(\bm{k})^{e-s} &=&\hbar v_F \left(\bm k+\frac{\beta^{e-s}}{2b}\pmb{\mathcal{A}}^{e-s} \right)\cdot\bm\sigma + \frac{\Delta}{2}\sigma_z,	\label{Eq:Hamiltonian-es}\\
\mathcal{H}_{K}(\bm{k})^{e-l} &=&\hbar v_F \left(\bm k+\frac{\beta^{e-l}}{b^2}\pmb{\mathcal{A}}^{e-l} \right)\cdot\bm\sigma + \frac{\Delta}{2}\sigma_z,	\label{Eq:Hamiltonian-el}
\end{eqnarray}
where $\beta^{e-s}$, $\beta^{e-l}$ are dimensionless coupling constants and $b=a_0/\sqrt{3}$ is the nearest-neighbor distance between carbon atoms, $a_0$ being the lattice constant. Since the effect of the gauge fields is to shift the Dirac cone of an amount proportional to the dimensionless coupling constants, $\beta^{e-s}$ and $\beta^{e-l}$ can be estimated from first principles from the Dirac-cone shift in reciprocal space induced by a lattice strain or distortion.

\section{Piezoelectricity} \label{Sec:piezo}

We first consider piezoelectricity in gapped graphene, wondering if large internal polarizations may arise
in response to a strain (cell) deformation. Such strain-induced polarization would affect  the electron-phonon coupling between charge carriers and acoustic phonons with foreseeable consequences on graphene resistivity\cite{PhysRevB.90.125414}.
As previously noticed, for the static approximation to hold the characteristic vibrational energy should be smaller than the bandgap, a condition that can be easily fulfilled for acoustic phonons involved in the electron scattering. 
Indeed, the maximum energy of acoustic phonons scattering with conduction electrons in graphene can be estimated as $\epsilon_{LA}\simeq v_s^{LA} 2k_F $, where $v_s^{LA}=21.4~$km/s  is the longitudinal sound velocity and $k_F=\epsilon_F/v_F$, $\epsilon_F$ being the Fermi energy and $v_F=10^{3}~$km/s the Fermi velocity\cite{PhysRevB.90.125414}. Usually, $\vert \epsilon_F\vert< 200~$meV in transport devices, resulting in a maximum acoustic phonon energy $\epsilon_{LA}\lesssim 4~$meV, which is much smaller than the typical bandgap measured in epitaxial graphene.
In this section, therefore, we estimate the polarization induced by acoustic phonons in the long wavelength limit and in the static (adiabatic) approximation, i.e., the piezoelectric constant.

As previously discussed, gapped distorted graphene with $\sqrt{3}\times\sqrt{3}\ \text{R30}^\circ$ reconstruction cannot be piezoelectric, whereas disproportionated graphene and gated bilayer graphene both display piezoelectricity. In the following we will mainly focus on the in-plane piezoelectric response, for which sizeable values -- ranging from $\sim 1.4\times 10^{-10}$ to $\sim 5.5\times 10^{-10}~C/m$ -- have been recently predicted in inhomogeneous 2D hexagonal crystals, such as h-BN and transition-metal dichalcogenide monolayers~\cite{DuerlooJPCL2012,Droth_PRB2016,Rostami_njp2d2018, piezo2d_review}.
In Ref. \onlinecite{Rostami_njp2d2018}, the piezoelectric constant of h-BN has been derived from model Eq. (\ref{Eq:Hamiltonian-es}) in  the framework of the modern theory of polarization~\cite{PhysRevB.48.4442,RevModPhys.66.899}.
Evaluating $\bm P$ within the Berry-phase formalism (see App. \ref{App:berry}), the piezoelectric constant can be related to the valley Chern number $C_{valley}=\sum_\tau \tau\,C_\tau$, where $C_\tau=\tau\sgn{\Delta}/2$ is the usual Chern number at a given valley labeled by $\tau$, where $\tau=\pm 1$ at $K$,$K'$ points, respectively.  Notice that the additional $\tau$ factor in the definition of valley Chern number guarantees that the contributions from inequivalent valleys sum up, leading to $C_{valley}=\sgn{\Delta}$, in contrast to the total Chern number of gapped graphene, that vanishes because the contributions from $K, K'$ points cancel out each other. 
Finally, the piezoelectric constant for gapped graphene monolayer in the Dirac-cone approximation is:
\begin{equation}\label{Eq:BP_piezo}
e_{222} = \beta^{e-s}\,\frac{e}{2\pi b} \,C_{valley} \equiv \beta^{e-s}\,\frac{e}{2\pi b} \,\sgn{\Delta},
\end{equation}
which differs by a factor 2 from the analytic expression previously given for h-BN\cite{Rostami_njp2d2018}. As discussed in App. \ref{App:berry}, this discrepancy is due to the fact that the complex conjugate contribution to the derivative of the Berry-phase polarization has been overlooked in Ref. \onlinecite{Rostami_njp2d2018}.

The piezoelectric constant, therefore, turns out to be independent on the Fermi velocity and the gap amplitude, while it depends only on the bond length $b$ and on the electron-strain coupling constant $\beta^{e-s}$. 
In graphene, such dimensionless coupling constant  can be derived from the Dirac-cone shift under a strain deformation calculated in Ref. \onlinecite{PhysRevB.90.125414} for both  clamped and relaxed ions\cite{PhysRevB.90.125414}, yielding $\beta^{e-s}_{\,c.i.}=3.37$ and $\beta^{e-s}_{\,r.i.}=2.66$, respectively\footnote{The electron-strain coupling constants $\beta_A$ calculated in Ref. \onlinecite{PhysRevB.90.125414} have the dimensions of energy and depend on the computational approach, with relative variations of ~15\% between DFT and GW values. The dimensionless coupling constant is derived as $\beta^{e-s}=2\sqrt{2}b\beta_A/\hbar v_F$; since also the Fermi velocity evaluated in DFT and GW shows deviations that are comparable to those found for $\beta_A$, their ratio (hence, the dimensionless coupling constant $\beta^{e-s}$) is substantially independent on the computational approach.}; these values compare well with those estimated for h-BN monolayer~\cite{Droth_PRB2016}, being $\beta^{e-s}_{\,c.i.}$(h-BN)$=3.3$ and $\beta^{e-s}_{\,r.i.}$(h-BN)$= 2.3$.
Since the Berry-phase expression of the piezoelectric coefficient \emph{does not depend on the bandgap}, disproportionated graphene is expected to display piezoelectric response comparable with  that  of the wide-gap h-BN monolayer as soon as the sublattice symmetry breaking is switched on. In fact, the relaxed-ion piezoelectric constant of gapped graphene monolayer can be estimated from Eq.\ (\ref{Eq:BP_piezo}), yielding $e_{222}=4.73\times 10^{-10}~C/m$, which is one order of magnitude larger than in chemically doped graphene (with a predicted piezoelectric constant $e_{222}\sim 6.3\times10^{-11} ~C/m$ )~\cite{Ong2013}. To put the estimated magnitude into context, an approximate 3D piezoelectric coefficient can be evaluated by dividing the 2D value of $e_{222}$ by the graphite interlayer spacing (3.35 \AA), yielding $e_{222, 3D}\sim 1.4~C/m^2$, which compares well with popular bulk piezoelectric materials such as ZnO or $\alpha-$quartz\cite{ZnOpiezo,quartzpiezo}.

\begin{figure}[h]
 \centering
 \includegraphics[scale=0.65]{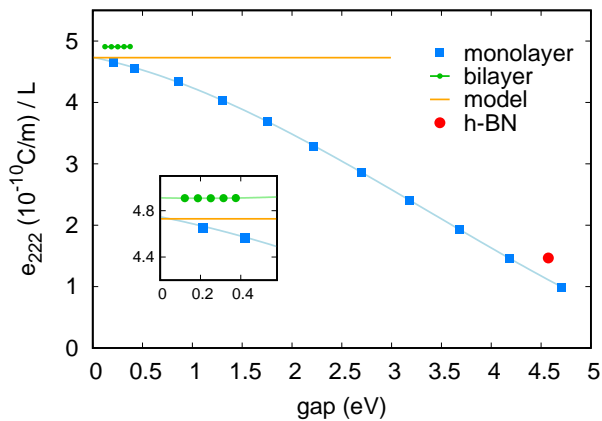}
 \caption{Relaxed-ion per-layer piezoelectric coefficient in disproportionated graphene and gated bilayer graphene as a function of the gap. The piezoelectric constant of h-BN monolayer, also shown, is larger that that of fully disproportionated graphene. This is due to the use of a slightly different lattice parameter in the two cases (as discussed in Sec. \ref{Sec:methods}a). The constant (orange) line represents the result of the analytical model Eq.\ (\ref{Eq:BP_piezo}). The inset is a zoom on the small gap region. Continuous lines are polynomial fits.}	\label{Fig:piezo}
\end{figure}

In order to further verify this prediction, we calculated from first principles the relaxed-ion piezoelectric coefficient of both disproportionated graphene and gated bilayer graphene as a function of the bandgap. 
The calculated per-layer piezoelectric coefficient is shown in  Fig.~\ref{Fig:piezo}, and compared  with the {\sl ab initio} estimate of the piezoelectric coefficient for h-BN~\cite{DuerlooJPCL2012}.
Our analytical result is well confirmed, even though $e_{222}$ is significantly suppressed as the bandgap increases in monolayer graphene, questioning the validity of the Dirac-cone model and/or gauge-field approximation for the electron-strain coupling for large values of the gap. 
Nonetheless, in the range of experimentally accessible values for the gap, $\Delta\lesssim 0.5~eV$, the per-layer piezoelectric coefficient $e_{222}$ of both gapped disproportionated graphene and gated bilayer graphene is substantially independent on the gap width. In addition, for gapped mono- and bilayer graphene the piezoelectric coefficient per layer is three times larger than that of the polar h-BN monolayer ($e_{222}=1.38\times 10^{-10}~C/m$)\cite{DuerlooJPCL2012} and $\sim$30$\%$ larger than that of the polar semiconducting MoS$_2$ monolayer ($e_{222}=2.5-4\,\times 10^{-10}~C/m$)\cite{piezo2d_review}.
Since gated bilayer graphene can be experimentally realized~\cite{NMat7.2007}, our prediction suggests a viable alternative for engineering large, measurable piezoelectricity in graphene that could be easier than the already proposed mechanisms based on chemical doping or asymmetric lattice defects\cite{SharmaAPL2012,Ong2012,Ong2013}. In addition, our findings suggest that the scattering by piezoelectric acoustic phonons can be sizable in gapped bilayer graphene, and it should be further explored its possible relevance for electric transport and charge-carrier relaxation.

\section{Born effective charges}\label{Sec:BEC}

We now come to analyze the evolution of the Born effective charge tensor of graphene induced by each symmetry-breaking scheme previously introduced. 

\begin{figure}[H]
\centering
\includegraphics[scale=0.65]{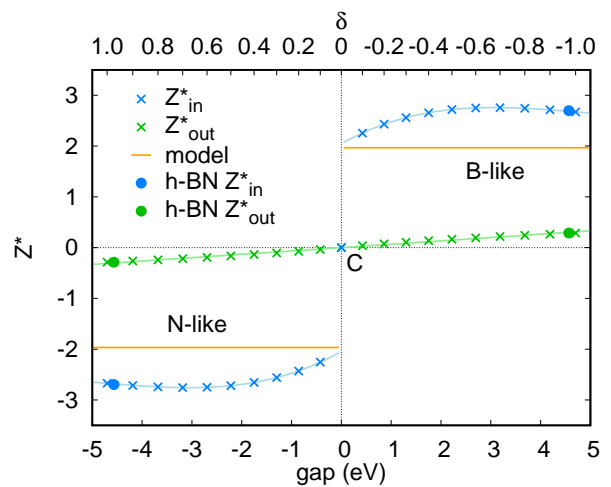}
\caption{In-plane (blue) and out-of-plane (green) components of the Born
effective charge in \emph{disproportionated} graphene of the atom with
nuclear charge $Z_{C2}=6+\delta$  as a function of the gap and of the
parameter $\delta$. Positive, negative and zero gap correspond to
B-like, N-like, C atom respectively (the sign of the gap is only a
graphical artifice). Continuous lines are polynomial fits. Filled
circles correspond to the Born tensor components of h-BN. The constant
line (orange) represents the result of the analytical model Eq.\
(\ref{Eq:berryBEC}).}	\label{Fig:disprop_charge}
\end{figure}

\paragraph*{a. Disproportionated graphene.}

The calculated in-plane and out-of-plane components of the Born effective charge are shown in Fig.~\ref{Fig:disprop_charge} as a function of the gap for monolayer graphene with inequivalent sublattices.
Only the effective charge of a single atom is shown since the {\color{black}other can be trivially obtained by imposing the ASR, as discussed in Sec. \ref{Sec:sym_break}}. The out-of-plane component of the Born tensor is found to increase linearly with the bandgap amplitude, vanishing when the equivalence between sublattices is restored, as expected. This can be rationalized by the fact that out-of-plane distortions do not alter - at the first order in the atomic displacement - the bond length, implying that the variation of charge density is mainly due to the rigid movement of the electrons together with the vibrating atom. 
On the contrary, an in-plane atomic displacement affects the bond length at linear order and thus radically modifies the electronic charge density, leading to a significant polarization response. Moreover, Fig.~\ref{Fig:disprop_charge} shows that the in-plane component of the Born tensor tends to a finite value when approaching the small gap limit -- even if the Born effective charges of pristine graphene vanish, as discussed in Sec.~\ref{Sec:sym_break} --, showing a step discontinuity as the gap changes sign.

\begin{figure}[H]
 \centering
 \includegraphics[scale=0.65]{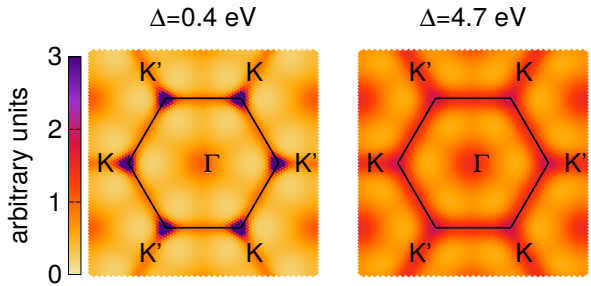}
 \caption{Reciprocal-space contributions to the in-plane component of the Born tensor in \emph{disproportionated} graphene for $\Delta=0.4$  eV ($\delta=0.1$) and $\Delta=4.7$ eV ($\delta=1.0$). The high symmetry points $\Gamma$, K and $\text{K}'$ of the reciprocal lattice are highlighted. This representation corresponds to the sum $Z_{xx}^\ast(\omega=0)+Z_{yy}^\ast(\omega=0)$ from Eq.\ \ref{Eq:dynamic_charge}.}	\label{Fig:kresolved}
\end{figure}

Similarly to piezoelectricity, this result seems to suggest that an arbitrarily small gap in disproportionated graphene is sufficient to have sizable Born effective charges, even comparable with the ones of h-BN monolayer. Further insight can be achieved by evaluating the Born effective charges from the Dirac-like model Eq.\  (\ref{Eq:Hamiltonian-el}). The validity of the Dirac approximation in this case is confirmed by Fig.~\ref{Fig:kresolved}, in which  we show the contributions of the interband transitions to the  Born effective charge tensor, calculated in DFPT using Eq.\ (\ref{Eq:dynamic_charge}) in the static limit ($\omega = 0$), as a function of the $\bm k$ index in the summation.
As one can see, for disproportionated graphene the most important contributions occur around the Dirac points $K$ and $K'$. The peaks at Dirac points get narrower as the bandgap is smaller, suggesting that the conic approximation is more accurate for small values of the gap.
The Born effective charges can then be evaluated from the Berry-phase polarization induced by the atomic displacements in model Eq.\ (\ref{Eq:Hamiltonian-el}), in close analogy with the calculation of piezoelectric coefficients (see App. \ref{App:berry}). It turns out that the in-plane Born effective charge also is related to the valley Chern number $\mathcal{C}_{valley}$, being:
\begin{equation}\label{Eq:berryBEC}
Z_{\parallel}=\beta^{e-l}\,\frac{3\sqrt{3}}{2\pi}\,C_{valley} \equiv \beta^{e-l}\,\frac{3\sqrt{3}}{2\pi}\sgn{\Delta}.
\end{equation}
Within this picture, the step discontinuity at zero gap corresponds to the change of sign of the sublattice potential, coinciding with the transition between the insulating and metallic states. The dimensionless coupling constant $\beta^{e-l}$ can be estimated from first principles as the ratio between the shifting of the Dirac cone, due to the atomic displacement $\mathbf{u}$, and the atomic displacement itself. We obtained $\beta^{e-l}=2.38$, which is consistent with the value $b^2\sqrt{\langle D_{\bm\Gamma}^2\rangle}/(\hbar v_F)=2.47$ given in Ref.~\citenum{Pisana2007}. Plugging this value in Eq.\ (\ref{Eq:berryBEC}), we find $Z_\parallel=1.97$, in very good agreement with the numerical results shown in  Fig.~\ref{Fig:disprop_charge}, especially for small values of the bandgap, where the Dirac-cone approximation is expected to hold.

Therefore, both our numerical and analytical results show that the in-plane Born effective charge - at least in the static approximation - is substantially independent on the bandgap for disproportionated graphene: a perturbation that breaks the equivalence between Carbon atoms induces an internal polarization, whose strength does not depend neither on the bandgap $\Delta$ nor on the Fermi velocity $v_F$ \emph{but only on the electron-phonon coupling constant $\beta^{e-l}$}, analogously to the piezoelectric coefficient dependence on the electron-strain coupling constant $\beta^{e-s}$.

\paragraph*{b. Distorted graphene.}

\begin{figure}[h]
\centering
\begin{tikzpicture}[node distance = 11mm, inner sep = 0pt]
\node (n0)  {\includegraphics[scale=0.65]{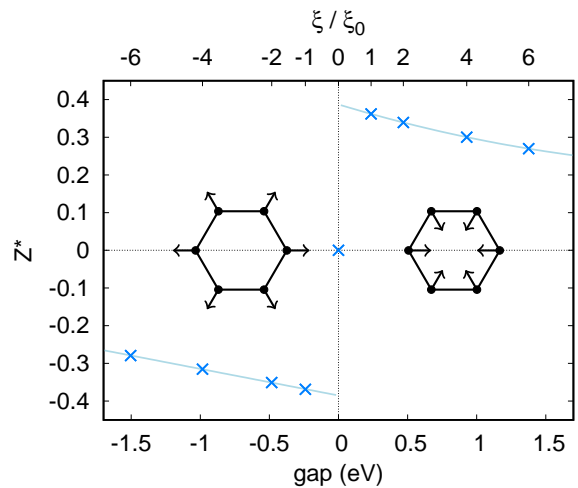}};
 \draw[thick] (-0.8,0) +(0:0.6cm) -- +(60:0.6cm) -- +(120:0.6cm) -- +(180:0.6cm) -- +(240:0.6cm) -- +(300:0.6cm) -- cycle;
 \node[inner sep=1pt,circle,draw,fill] at ($ (-0.8,0) + (0:0.6cm) $) {};
 \node[inner sep=1pt,circle,draw,fill] at ($ (-0.8,0) + (60:0.6cm) $) {};
 \node[inner sep=1pt,circle,draw,fill] at ($ (-0.8,0) + (120:0.6cm) $) {};
 \node[inner sep=1pt,circle,draw,fill] at ($ (-0.8,0) + (180:0.6cm) $) {};
 \node[inner sep=1pt,circle,draw,fill] at ($ (-0.8,0) + (240:0.6cm) $) {};
 \node[inner sep=1pt,circle,draw,fill] at ($ (-0.8,0) + (300:0.6cm) $) {};
 \draw [->,thick] ($ (-0.8,0) + (0:0.6cm) $) -- ($ (-0.8,0) + (0:0.9cm) $);
 \draw [->,thick] ($ (-0.8,0) + (60:0.6cm) $) -- ($ (-0.8,0) + (60:0.9cm) $);
 \draw [->,thick] ($ (-0.8,0) + (120:0.6cm) $) -- ($ (-0.8,0) + (120:0.9cm) $);
 \draw [->,thick] ($ (-0.8,0) + (180:0.6cm) $) -- ($ (-0.8,0) + (180:0.9cm) $);
 \draw [->,thick] ($ (-0.8,0) + (240:0.6cm) $) -- ($ (-0.8,0) + (240:0.9cm) $);
 \draw [->,thick] ($ (-0.8,0) + (300:0.6cm) $) -- ($ (-0.8,0) + (300:0.9cm) $);
 \draw[thick] (2,0) +(0:0.6cm) -- +(60:0.6cm) -- +(120:0.6cm) -- +(180:0.6cm) -- +(240:0.6cm) -- +(300:0.6cm) -- cycle;
 \node[inner sep=1pt,circle,draw,fill] at ($ (2,0) + (0:0.6cm) $) {};
 \node[inner sep=1pt,circle,draw,fill] at ($ (2,0) + (60:0.6cm) $) {};
 \node[inner sep=1pt,circle,draw,fill] at ($ (2,0) + (120:0.6cm) $) {};
 \node[inner sep=1pt,circle,draw,fill] at ($ (2,0) + (180:0.6cm) $) {};
 \node[inner sep=1pt,circle,draw,fill] at ($ (2,0) + (240:0.6cm) $) {};
 \node[inner sep=1pt,circle,draw,fill] at ($ (2,0) + (300:0.6cm) $) {};
 \draw [->,thick] ($ (2,0) + (0:0.6cm) $) -- ($ (2,0) + (0:0.3cm) $);
 \draw [->,thick] ($ (2,0) + (60:0.6cm) $) -- ($ (2,0) + (60:0.3cm) $);
 \draw [->,thick] ($ (2,0) + (120:0.6cm) $) -- ($ (2,0) + (120:0.3cm) $);
 \draw [->,thick] ($ (2,0) + (180:0.6cm) $) -- ($ (2,0) + (180:0.3cm) $);
 \draw [->,thick] ($ (2,0) + (240:0.6cm) $) -- ($ (2,0) + (240:0.3cm) $);
 \draw [->,thick] ($ (2,0) + (300:0.6cm) $) -- ($ (2,0) + (300:0.3cm) $);
\end{tikzpicture}
\caption{Radial component of the Born effective charge in distorted graphene as a function of the gap and the atomic displacement $\xi$ given in units of $\xi_0=8.875\times10^{-3}$~\AA. The sign of the gap  is the same of $\xi$ which is positive for the narrowing and negative for the enlargement of the hexagon. Continuous lines are polynomial fits.}		\label{Fig:hexag_charge}
\end{figure}
As previously discussed, the Born effective charge tensor of distorted graphene is diagonal in the radial-tangential coordinate system and the two components must have opposite sign as required by the ASR. 
In Fig.~\ref{Fig:hexag_charge} we report the radial component of the Born tensor as a function of the gap and the atomic displacement $\xi$ expressed in units of $\xi_0$ (see Sec. \ref{Sec:methods}).
As for disproportionated graphene, the in-plane Born effective charge becomes suddenly different from zero as soon as the perturbation is switched on, that is, in this case, as soon as the atoms are displaced from the original position.
However, for distorted graphene the magnitude of the in-plane Born effective charge is rather reduced compared to disproportionated graphene, being roughly one order of magnitude smaller.
We also notice that the Born effective charge behaves slightly asymmetrically with respect to the atomic displacement.
Such behavior is not surprising, since the distortion does not preserve the translational symmetry of graphene and a positive atomic displacement is not equivalent to a negative one.

\paragraph*{c. Gated bilayer graphene.}

\begin{figure}[ht]
\centering\normalsize
\begin{tikzpicture}[node distance = 11mm, inner sep = 0pt]
\node (n0)  {\includegraphics[scale=0.65]{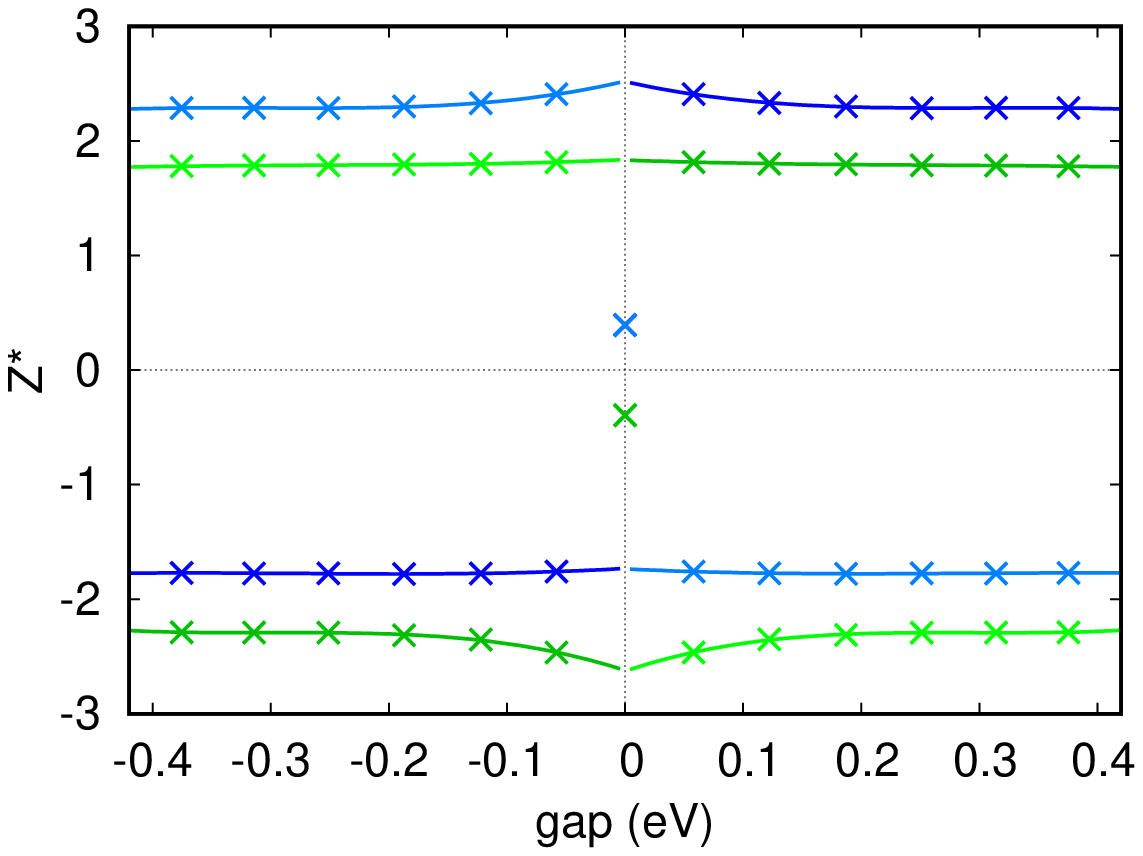}};
 \draw[thick] (-1.5,-.08) -- (-1,-.08);
 \draw[thick] [densely dashed] (-1,-.08) -- (-1,.72);
 \draw[thick] (-1,.72) -- (-0.5,.72);
 \node[inner sep=1.5pt,circle,draw,fill,green,label={[label distance=1mm]270: 3}] at (-1.5,-.08) {};
 \node[inner sep=1.5pt,circle,draw,fill,blue,label={[label distance=1mm]270: 1}] at (-1,-.08) {};
 \node[inner sep=1.5pt,circle,draw,fill,cyan,label={[label distance=0.8mm]90: 2}] at (-1,.72) {};
 \node[inner sep=1.5pt,circle,draw,fill,black!30!green,label={[label distance=0.8mm]90: 4}] at (-0.5,.72) {};
 \draw [thick,->] (2.8,-0.08) -- (2.8,0.72) node[midway, right=+1mm] {\textbf{E}};
 \draw [thick,<-] (-2.5,-0.08) -- (-2.5,0.72) node[midway, right=+1mm] {\textbf{E}};
\end{tikzpicture}
\caption{In-plane component of the Born tensors in gated bilayer graphene as a function of the gap at K.
The sign of the gap corresponds to the direction of the electric field with respect to the z axis. The colors relate the curves with the corresponding atom. For zero electric field the Born effective charges $Z_1^\ast=Z_2^\ast=-Z_3^\ast=-Z_4^\ast=0.394$ were computed as the average between the right and left limit. Continuous lines are polynomial fits.}
\label{Fig:bilayer_charge}
\end{figure}

\begin{figure}[ht]
\centering\normalsize
\begin{tikzpicture}[node distance = 11mm, inner sep = 0pt]
\node (n0)  {\includegraphics[scale=0.65]{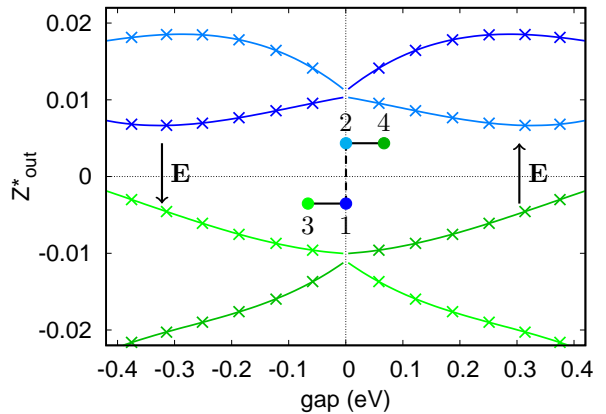}};
 \draw[thick] (-0.08,-.08) -- (0.42,-.08);
 \draw[thick] [densely dashed] (0.42,-.08) -- (0.42,.72);
 \draw[thick] (0.42,.72) -- (0.92,.72);
 \node[inner sep=1.5pt,circle,draw,fill,green,label={[label distance=1mm]270: 3}] at (-0.08,-.08) {};
 \node[inner sep=1.5pt,circle,draw,fill,blue,label={[label distance=1mm]270: 1}] at (0.42,-.08) {};
 \node[inner sep=1.5pt,circle,draw,fill,cyan,label={[label distance=0.8mm]90: 2}] at (0.42,.72) {};
 \node[inner sep=1.5pt,circle,draw,fill,black!30!green,label={[label distance=0.8mm]90: 4}] at (0.92,.72) {};
 \draw [thick,->] (2.7,-0.08) -- (2.7,0.72) node[midway, right=+1mm] {\textbf{E}};
 \draw [thick,<-] (-2.0,-0.08) -- (-2.0,0.72) node[midway, right=+1mm] {\textbf{E}};
\end{tikzpicture}
\caption{Out-of-plane component of the Born tensors in gated bilayer graphene as a function of the gap at K.
The sign of the gap corresponds to the direction of the electric field with respect to the z axis. The colors relate the curves with the corresponding atom. Continuous lines are polynomial fits.}
\label{Fig:bilayer_charge_out}
\end{figure}

In Fig.~\ref{Fig:bilayer_charge} we show the evolution of the in-plane component of the Born effective charge tensors in gated bilayer graphene as a function of the gate-induced bandgap. We observe that the Born effective charges of gapped bilayer exhibit analogous features as observed in the monolayer: {\it i)} they are almost independent on the gap, and thus on the intensity of the electric field which induces the gap in the bilayer and {\it ii)}
they get sizable as soon as the electric field, and thus the gap, is different from zero. The numerical value $Z_\parallel\approx 2$ is even similar to the one found in small gap region of disproportionated graphene, again in close analogy with the predicted per-layer piezoelectric coefficient discussed in the Sec.~\ref{Sec:piezo}. On the other hand, the out-of-plane component of the effective charge tensor is smaller than 1~\% of the in-plane component, as shown in Fig. \ref{Fig:bilayer_charge_out}.
At zero field, the symmetry arguments discussed in Section~\ref{Sec:sym_break} allow for a finite albeit small  value of the effective charge for bilayer graphene. In Fig.~\ref{Fig:bilayer_charge} we estimate $Z_\parallel\sim 0.39$ as the average between the right and left limit of the calculated $Z^{I}_{\parallel}$, which is found to fulfill the charge-neutrality sum rule $Z^{C1}_{\parallel}=Z^{C2}_{\parallel}=-Z^{C3}_{\parallel}=-Z^{C4}_{\parallel}$. %

All symmetry-breaking mechanisms discussed above show similar features. Beside inducing an opening of the bandgap, they invariably trigger finite and sizable (in-plane) Born effective charges, which weakly depend on the gap value. Additionally, both for gapped monolayer graphene and gated bilayer graphene the effective charges are comparable to those of h-BN, showing a remarkable universal behavior. At the same time, in all the considered cases this universality comes along with a discontinuous behavior at zero gap. This effect is a signature of the failure of the static approximation when the gap becomes comparable to the phonon frequency. As we shall see in the next section, this \emph{paradox} is solved once that the dynamical (non-adiabatic) effects are taken into account, allowing one to recover the expected vanishing of the IR phonon strength for pristine graphene.

\section{Non-adiabatic effective charges in disproportionated graphene}   \label{Sec:model}

We will focus here on the dynamical generalization of Born effective charges in disproportionated graphene, where 
they can be explicitly calculated within time-dependent DFPT using Eq.\ (\ref{Eq:dynamic_charge}) for the Dirac-like model Eq.\ (\ref{Eq:Hamiltonian-el}). The velocity and electron-phonon operators entering in Eq.\ (\ref{Eq:dynamic_charge}) can be readily evaluated in the Dirac-cone approximation, being:
\begin{equation}\label{defvertices}
v_\alpha=v_F\sigma_\alpha, \quad \frac{\partial H_{\bf K}}{\partial u_\mu^A}=\hbar v_F \frac{\beta^{e-l}}{b^2} (\bm\sigma\times\mathbf{\hat z})_\mu.
\end{equation}
The dynamical effective charge can thus be written in a compact form as
\begin{equation}
 Z^{A,el}_{\alpha\mu}(\omega) = \frac{2e}{N_k} \sum_\mathbf{k} \sum_{i,j} C_{\alpha\mu,\mathbf{k}}^{ij} \Pi_k^{ij}(\omega)	\label{Eq:dynamic_charge_graphene}
\end{equation}
where
\begin{equation}
C_{\alpha\mu,\mathbf{k}}^{ij} = \langle u_{\mathbf{k}i} | \frac{i\hbar v_F\sigma_\alpha}{\epsilon_{\mathbf{k}i}-\epsilon_{\mathbf{k}j}} | u_{\mathbf{k}j} \rangle \langle  u_{\mathbf{k}j} | \hbar v_F \frac{\beta^{e-l}}{b^2} (\bm\sigma\times\mathbf{\hat z})_\mu | u_{\mathbf{k}i}\rangle
\end{equation}
and
\begin{equation}\label{pijk}
\Pi_k^{ij}(\omega) = \frac{\theta(\epsilon_F-\epsilon_{\mathbf{k}i})-\theta(\epsilon_F-\epsilon_{\mathbf{k}j})}{\epsilon_{\mathbf{k}i}-\epsilon_{\mathbf{k}j}+\hbar\omega+i\eta}.
\end{equation}
Here, $\epsilon_F$ is the Fermi energy, $\theta$ is the Heaviside step-function and $\eta$ is an infinitesimal positive number needed to define the retarded response function. However, to simulate the effect of a finite electronic lifetime, we will explicitly derive in the following the expression for the effective charge by assuming that $\eta$ can acquire a finite value. Notice that, in the presence of doping, the excess electronic density  per unit area with respect to the charge-neutrality condition can be easily computed for gapped graphene in the Dirac-cone approximation, being:
\begin{equation}
 n=2	 \int_0^{k_F} \frac{kdk}{2\pi} = \frac{1}{\pi(\hbar v_F)^2} \left(\epsilon_F^2-\Delta^2/4\right)\theta\left(\epsilon_F-\Delta/2\right).
\end{equation}
Using Equations~\ref{Eq:eigenthings} and \ref{Eq:coefficients} and replacing the summation over ${\bm k}$ with the integral $\Omega\int kdk/2\pi\int d\theta/2\pi$ we get
\begin{equation}
 Z^{A,el}_{\alpha\alpha}(\omega) = 2eN_v\Omega \int_0^{\bar k} \frac{kdk}{2\pi} \sum_{\zeta=\pm1} \frac{-\langle C_{\alpha\alpha,\mathbf{k}}^{\pi\pi^\ast} \rangle_\theta}{2E+\zeta(\hbar\omega+i\eta)}, 	\label{Eq:integral_charge}
\end{equation}
where $N_v=2$ is the valley degeneracy and $\bar k$ is a momentum cut-off for the linear conic approximation, whose exact value is irrelevant since the integral (\ref{Eq:integral_charge}) is convergent at large momenta. Finally
\begin{equation}
\langle C_{\alpha\alpha,\mathbf{k}}^{\pi\pi^\ast} \rangle_\theta=\langle C_{\alpha\alpha,\mathbf{k}}^{\pi^\ast\pi} \rangle_\theta=-(\hbar v_F)^2 \frac{\beta^{e-l}}{b^2} \frac{\Delta}{4E^2}
\end{equation}
is the average over $\theta$ of $C_{\alpha\alpha,\mathbf{k}}^{\pi\pi^\ast}$.
Performing the integral over the energy and separating real and imaginary part of the effective charge tensor, we obtain
\begin{align}
\label{rezaa}
 Re[Z^{A,el}_{\alpha\alpha}(\omega)]=& \tilde Z \frac{\Delta\hbar\omega}{(\hbar\omega)^2+\eta^2} \left[\frac{\eta}{\hbar\omega} \tilde\theta(m,\omega) + \tilde\xi(m,\omega) \right], \\
 \label{imzaa}
 Im[Z^{A,el}_{\alpha\alpha}(\omega)]=& \tilde Z \frac{\Delta\hbar\omega}{(\hbar\omega)^2+\eta^2}  \left[\tilde\theta(m,\omega) - \frac{\eta}{\hbar\omega} \tilde\xi(m,\omega)\right],
\end{align}
where $\tilde Z$ is the static effective charge calculated by taking $\omega = 0$ in Eq.\ (\ref{Eq:integral_charge}) and the functions $\tilde\theta(m,\omega)$ and $\tilde\xi(m,\omega)$ are given by
\begin{align}
 \tilde\theta(m,\omega)&=\pi-\tan^{-1}\frac{2m+\hbar\omega}{\eta}-\tan^{-1}\frac{2m-\hbar\omega}{\eta},	\label{Eq:theta-tilde} \\
\tilde\xi(m,\omega)&=\frac{1}{2}\log{\frac{(2m+\hbar\omega)^2+\eta^2}{(2m-\hbar\omega)^2+\eta^2}},		\label{Eq:csi-tilde}
\end{align}
with $m$ being the threshold energy for particle-hole excitations:
\begin{equation}\label{def:m}
m=\max(\Delta/2,\epsilon_F).
\end{equation}
We notice that, in the absence of doping ($n=0$), we recover the static Born effective charge evaluated from the Berry-phase polarization, as expected. This can be easily seen from Eq.\ (\ref{Eq:integral_charge}), since for $\omega=0$ the integral only depends on the dimensionless variable $v_F k/\Delta$, hence:
\begin{equation}
 \tilde Z=Z^{A,el}_{\alpha\alpha}(\omega=0)= \beta^{e-l} \frac{3\sqrt{3}}{2\pi}\sgn{\Delta},	\label{Eq:static_charge_conic} 
\end{equation}
which coincides precisely with Eq. (\ref{Eq:berryBEC}), so that $\tilde Z=1.97$.
\begin{figure}[H]
 \centering
 \includegraphics[scale=0.65]{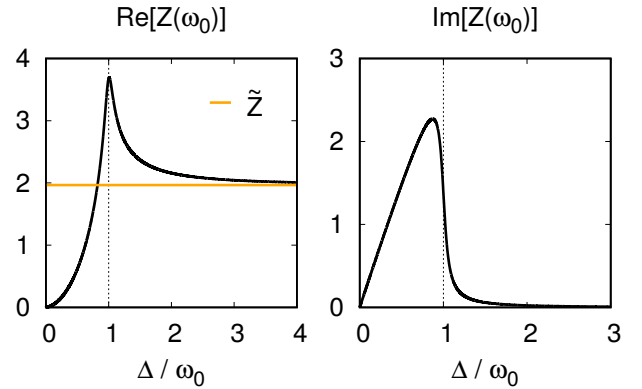}
 \caption{Real and imaginary part of the dynamical effective charge $Z(\omega_0)$ given by Eqs. (\ref{rezaa}) and (\ref{imzaa}), respectively, as a function of the ratio $\Delta/\hbar \omega_0$. Here to mimic the finite electron lifetime we used a broadening $\eta=0.01$ eV, which is considerably smaller than the phonon energy scale $\hbar \omega_0\simeq 0.2$ eV. The constant line (orange) corresponds to the static effective charge $\tilde Z=1.97$ evaluated via Eq.\ (\ref{Eq:berryBEC}) or, equivalently, Eq.\ (\ref{Eq:static_charge_conic}).}   \label{Fig:dynamical_charge}
\end{figure}

While the (adiabatic) Born effective charge of disproportionated graphene is independent on the bandgap, the frequency-dependent, non-adiabatic effective charge displays a strong dependence on it. In fact, the integral in Eq.\ (\ref{Eq:integral_charge}) diverges when $\Delta\sim \hbar\omega$. As we will discuss in the next section, this dependence dramatically affects the optical vibrational response, especially when the bandgap is comparable with characteristic phonon energies.
To highlight this effect in disproportionated graphene, we show  in Fig.~\ref{Fig:dynamical_charge} the real and imaginary parts of the non-adiabatic effective charge evaluated at the frequency of the graphene optical phonon ($\hbar\omega_0 = 1582~ $cm$^{-1} \simeq 0.2~eV$) as a function of the gap $\Delta$.
As expected for pristine graphene, $ Z(\omega_0)$ vanishes at zero gap, while for $|\Delta|\gg\hbar\omega_0$ the imaginary part of the effective charge vanishes, and $ Z(\omega_0)$ is a real number which tends to the static value
$\tilde Z$. The dependence on the bandgap clearly emerges close to the resonance condition $\Delta \sim \hbar\omega_0$, as the frequency-dependent effective charge is peaked at the phonon energy.
In the next section, we will show that
the enhancement of the effective charge close to the resonance is responsible for an intensification of the infrared spectral line, once the graphene bandgap is engineered and tuned to match the optical phonon energy, while its imaginary part determines the phonon peak shape and, specifically, the Fano asymmetry.

\begin{figure}[ht]
 \centering
 \includegraphics[scale=0.75]{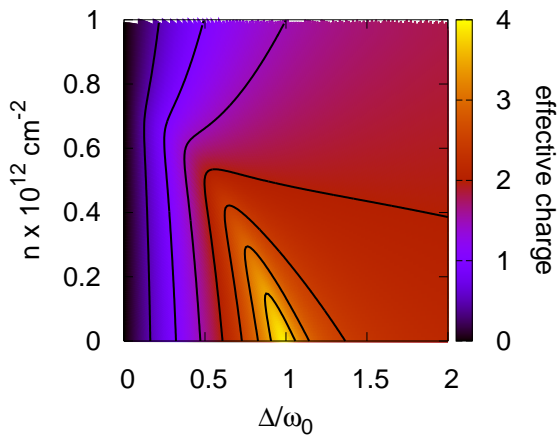}
 \caption{Absolute value of the dynamical effective charge at the phonon frequency as a function of the gap divided by $\hbar\omega_0$ and of the electron density $n$. Continuous lines correspond to semi-integer values of $|Z(\omega_0)|$.}	\label{Fig:3dplot}
\end{figure}

The non-adiabatic effective charge of disproportionated graphene also depends on the carrier density in the presence of doping. In Fig.~\ref{Fig:3dplot} we show a contour plot of the modulus of the effective charge at the phonon frequency $|Z(\omega_0)|$ as a function of the gap and doping. For $\Delta=0$ the effective charge vanishes whatever is the Fermi energy as expected for pristine graphene. For $n=0$ (undoped sample) the effective charge is peaked at the resonance and tends to the static limit for $\Delta>\hbar\omega_0$. In the intermediate region the effective charge forms a ridge whose height decreases with increasing doping.

\section{Optical Conductivity in disproportionated graphene}\label{Sec:optical}

{\color{black}It is well known that the effective charge controls the optical strength of the phonon response in the optical conductivity.} 
In 3D materials the frequency-dependent polarization and the current density 
 are given by $\mathbf{P}(\omega) = \bm\chi(\omega) \mathbf{E}(\omega)$
and $\mathbf{J}(\omega) = \bm\sigma(\omega) \mathbf{E}(\omega)$, respectively.
The two response functions, electrical susceptibility $\bm\chi(\omega)$ and optical conductivity $\bm\sigma(\omega)$, are 3$\times$3 Cartesian tensors obeying the relation
$\bm\sigma(\omega)=-i\omega\bm\chi(\omega)$.
The optical conductivity in SI units has dimensions of $\Omega^{-1}m^{-1}$, whereas in 2D crystals $\mathbf{J}(\omega)$ is given by the electric current per unit length and the optical conductivity has units of $\Omega^{-1}$.
The optical conductivity can be written as the sum of an electronic and an ionic contribution: $\bm\sigma(\omega)=\bm\sigma^{el}(\omega)+\bm\sigma^{ion}(\omega)$.
The former is due to the electronic current with ions kept fixed, while the latter arises from the nuclear motion as a consequence of the electric field.
Its  Cartesian components can be generally written as
\begin{equation}
 \sigma_{\alpha\beta}^{ion}(\omega) = -i\omega \frac{1}{\Omega} \sum_{s=4}^{3N} \frac{f_{s,\alpha}(\omega)f_{s,\beta}(\omega)}{\omega_s^2-(\omega+i\gamma_s/2)^2}	\label{Eq:ionic_susceptibility}
\end{equation}
where $s$ is phonon mode index (restricted to optical modes), $\omega_s$ is the phonon frequency,
$\gamma_s$ is the full width at half maximum of the phonon peak in the static limit and ${\bm f}_s(\omega)$ is the oscillator strength, defined as
\begin{equation}
 f_{s,\alpha}(\omega)= e \sum_{I,\mu} Z_{\alpha\mu}^I(\omega)e_{s,\mu}^I (M_I)^{-1/2}.	\label{Eq:oscillator_strength}
\end{equation}
Here $M_I$ is the mass of ion $I$ and ${\bm e}_s^I$ is the orthonormalized eigenvector of the dynamical matrix corresponding to the eigenvalue $\omega_s^2$, while $Z_{\alpha\mu}^I(\omega)$ is the frequency-dependent effective charge defined in  Eq.\ (\ref{Eq:dynamic_charge}).

From Eqs..\ (\ref{Eq:ionic_susceptibility}) and (\ref{Eq:oscillator_strength}) one immediately sees that  since the phonon peak is usually rather sharp its spectral weight is essentially controlled by the value of the dynamical effective charge tensor evaluated at the phonon frequency. As discussed below Eq.\ (\ref{Eq:dynamic_charge}), $Z^I_{\alpha\mu}(\omega_s)$ acquires a finite imaginary part in small-gap semiconductors, where the bandgap is smaller than $\omega_s$, allowing for  interband transitions at the energy scale of the phonon. In this regime the complex nature of the dynamical charge tensor explains also the  appearance of the Fano asymmetry of the optical spectra\cite{Cappelluti_prb10,Cappelluti_prb12}. Indeed, the infrared absorption -- proportional to the real part of the optical conductivity Eq.\ (\ref{Eq:ionic_susceptibility}) -- can be written in general for $\omega\simeq \omega_s$ as a Fano function\cite{PhysRev.124.1866}:
\begin{equation}
 Re[\sigma_{\alpha\beta}^{ion}(\omega)] = \sum_{s=4}^{3N} P_{s,\alpha\beta} \frac{q_\alpha q_\beta-1+(q_\alpha+q_\beta)z}{(1+q_\alpha q_\beta)(1+z^2)},	\label{Eq:fano_shape}
\end{equation}
where we introduced the variable $z=2[\omega-\omega_s]/\gamma_s$.  
The constant weight $P_{s,\alpha\beta}$, which represents the \textit{phonon strength} \cite{Cappelluti_prb12}, is defined as
\begin{align}
P_{s,\alpha\beta}=\frac{1}{\Omega\gamma_s}\{Re[f_{s,\alpha}(\omega_s)]Re[f_{s,\beta}(\omega_s)]+\nonumber \\ +Im[f_{s,\alpha}(\omega_s)]Im[f_{s,\beta}(\omega_s)]\}
\end{align}
and the Fano asymmetry parameter $q_\alpha$ is given by
\begin{equation}
\label{Eq:q}
q_\alpha=-\frac{Re[f_{s,\alpha}(\omega_s)]}{Im[f_{s,\alpha}(\omega_s)]}.
\end{equation}

\begin{figure}[h]
\centering
\includegraphics[scale=0.65]{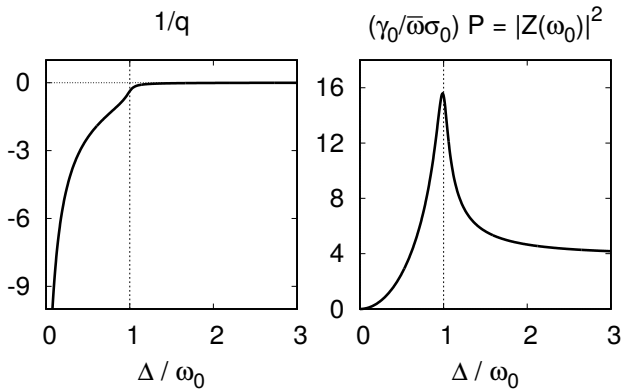}
\caption{(a) Inverse Fano asymmetry parameter $1/q$ and (b) phonon strength $P$ (in units of $\bar\omega\sigma_0/\gamma_0$) in undoped gapped graphene as a function of the gap value, as given by Eqs.\ (\ref{defq}) and (\ref{defp}). }	\label{Fig:parameters}
\end{figure}

\begin{figure}[h]
\centering
\includegraphics[scale=0.75]{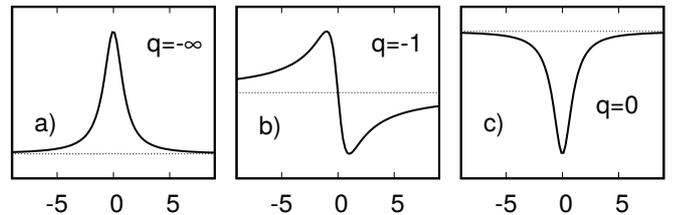}
\caption{ Sketch of the Fano shaped peak in different regimes: (a) positive symmetric Lorentzian peak; (b) highly antisymmetric Fano profile (q=-1); (c) negative phonon peak (q=0). The peak-to-valley distance is equal to 1 in all the cases.}	\label{Fig:Fano}
\end{figure}

In the specific case of gapped graphene, the results of the previous Section can be readily used to compute the ionic contribution (\ref{Eq:fano_shape}) of the $G$ mode. In this case the dynamical charge tensor is diagonal, and given by Eqs.\ (\ref{rezaa})-(\ref{imzaa}). Eq.\ (\ref{Eq:fano_shape}) then simplifies to 
\begin{equation}
 Re[\sigma^{ion}_{xx}(\omega\approx\omega_0)]=P \frac{q^2-1+2qz}{(1+q^2)(1+z^2)}	\label{Eq:conductivity_ion}
\end{equation}
where $\omega_0$ and $\gamma_0$ denote the frequency and linewidth of the $G$ mode, and 
\begin{eqnarray}
\label{defp}
P&=&\frac{2 e^2}{\Omega M_I}\frac{|Z_{xx}^{el}(\omega_0)|^2}{\gamma_0}=\bar \omega \sigma_0 \frac{|Z_{xx}^{el}(\omega_0)|^2}{\gamma_0} \\
\label{defq}
q&=&-Re[Z_{xx}^{el}(\omega_0)]/Im[Z_{xx}^{el}(\omega_0)]
\end{eqnarray}
where  the electronic strength has been expressed in terms of the so-called universal value $\sigma_0=4 e^2/\hbar$, which sets the scale of the electronic optical absorption in graphene\cite{2008IJMPB..22.2529P}, and $\hbar \bar \omega=8\hbar^2/M_I \Omega=0.53$ meV. 
To understand the effect of the gap opening in graphene we show explicitly in Fig.\ \ref{Fig:parameters} the $q$ and $P$ parameters as a function of $\Delta$. As discussed before,  when the bandgap is larger than the phonon frequency  the imaginary part of the effective charge vanishes, so that from Eq.\ (\ref{defq}) $q=-\infty$, and one recovers the usual symmetric Lorentzian profile. This can be seen in the first panel of Fig.~\ref{Fig:Fano}, where we show few illustrative cases of Eq.\ (\ref{Eq:conductivity_ion}) at different values of the $q$ for  a fixed value of $P=1$. As the bandgap progressively decreases the particle-hole continuum overlaps with the phonon frequency giving rise to a complex effective charge, which leads to a finite value of $q$ and an emerging asymmetric Fano profile [Fig.~\ref{Fig:Fano}(b)]. When the resonance condition $\omega_0\approx \Delta$ between electronic excitations and the phonon frequency is maximized the Fano parameter approaches $q=0$, and a negative peak appears [Fig.~\ref{Fig:Fano}(c)]. At the same time, in gapped graphene also the real part of the effective charge displays a non-trivial frequency dependence at gap values smaller than the phonon energy, as already discussed in the previous Section. This has a marked effect on the phonon strength (\ref{defp}) shown in Fig.\ \ref{Fig:parameters}, since it vanishes as expected when $\Delta \rightarrow 0$ but displays a strong enhancement in proximity of the resonance condition $\omega_0\approx \Delta$. 

To compare directly the ionic optical absorption  with the electronic one we estimated also the latter in the conic approximation. Its analytical expression has been already derived in Ref.~\citenum{PhysRevLett.96.256802}, 
including both intra- and inter-band contributions. In the limit of long electronic lifetime ($\eta\to 0$)
it can be readily shown that the real part of the interband contribution, the only one we are interested in, is given by (see Appendix~\ref{App:conductivity})
\begin{equation}
Re[\sigma^{el}_{xx}(\omega)]=\sigma_0\left(1+\frac{\Delta^2}{\hbar^2\omega^2}\right)\theta\left(\frac{\hbar\omega}{2}-m\right),	\label{Eq:conductivity_el}
\end{equation}
where $m$ is the excitation threshold defined in Eq.\ (\ref{def:m}) above. Finally, to compute the ionic part (\ref{Eq:conductivity_ion}) we also need an estimate of the phonon linewidth  $\gamma_0$. This is given by the sum of two different contributions $\gamma^{e-ph}+\gamma^{ph-ph}$~\cite{PhysRevLett.99.176802}.
The first one, $\gamma^{e-ph}$, is associated with the phonon decay into particle-hole pairs and can be explicitly computed in the conic approximation as (see Appendix~\ref{App:self-energy}):
\begin{equation}
 \gamma^{e-ph}=\bar \gamma \left( 1+\frac{\Delta^2}{\hbar^2\omega_0^2} \right) \theta\left(\frac{\hbar\omega_0}{2}-m\right),	\label{Eq:broadening}
\end{equation}
where $\bar \gamma=9\sqrt{3}\hbar\beta^2/4Ma^2\simeq1.26$ meV (10.2 cm$^{-1}$). 
The second one, representing the decay of one phonon into two phonons,  has been fixed to a small constant value $\gamma^{ph-ph}\simeq 0.25$ meV (2 cm$^{-1}$, see Fig.~1 of Ref.~\citenum{2008IJMPB..22.2529P}).
For finite electronic lifetime Eqs. (\ref{Eq:conductivity_el}), (\ref{Eq:broadening}) can be readily modified, in analogy with the general expressions for the dynamical effective charge given by Eqs.\ (\ref{rezaa})-(\ref{imzaa}), and the full expressions are given in Appendices~\ref{App:conductivity} and \ref{App:self-energy}, respectively.

\begin{figure}[H]
 \centering
 \includegraphics[scale=0.65]{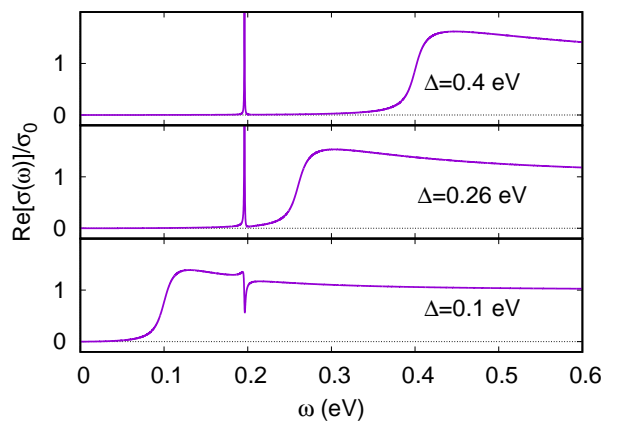}
 \caption{Real part of the total optical conductivity as a function of the frequency $\omega$ for different values of $\Delta$.}	\label{Fig:conductivity}
\end{figure}
The real part of the total optical conductivity  in absence of doping - sum of electronic and ionic contribution - is shown in Fig.~\ref{Fig:conductivity} as a function of the frequency $\omega$ for different values of the bandgap $\Delta$.
For $\Delta>\hbar\omega_0\simeq $0.2 meV the effective charge is real and well approximated by the static value $\tilde Z$ given by Eq.\ (\ref{Eq:static_charge_conic}). Since $q\rightarrow \infty$ very rapidly, already for $\Delta=0.4$~eV the phonon peak presents a Lorentzian shape, and it appears as a sharp and very intense peak inside the electronic gap. As the gap value decreases and approaches the $\omega_0$ value the dynamical effective charge gets further enhanced with respect to the static limit, and the phonon response acquires a progressively larger optical strength $P$. Finally, when $\Delta<\hbar\omega_0$ the ionic contribution overlaps with the electronic one, the effective charge becomes complex and the Fano parameter gets progressively smaller, leading to a pronounced Fano asymmetry which manifests with a marked negative peak. 
The same analysis can be done for doped graphene with substantially no conceptual differences.

\begin{figure}
 \centering
 \includegraphics[scale=0.65]{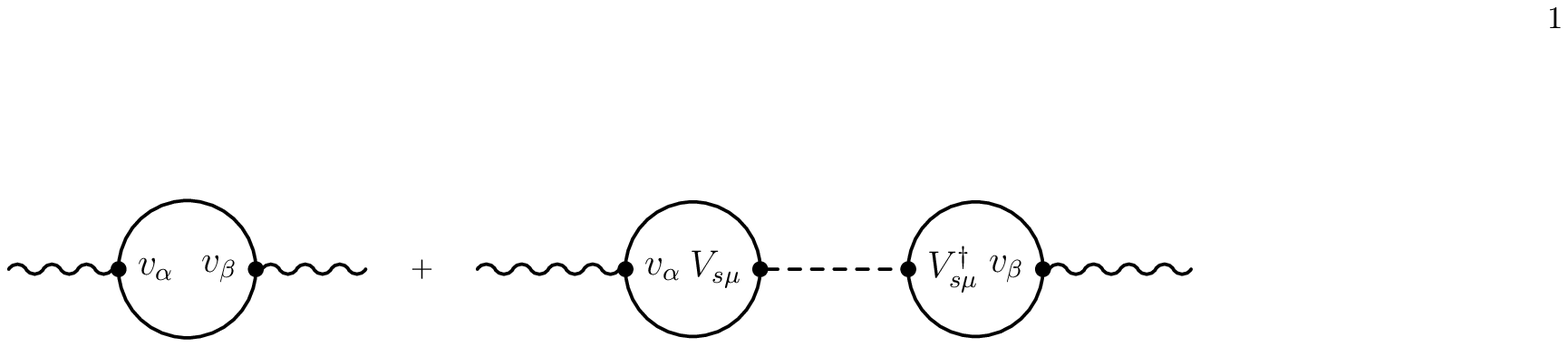}
 \caption{Feynman diagrams representing the calculation of the optical conductivity in the standard field-theory formalism, along the lines of the approach used in Ref.\ \citenum{Cappelluti_prb10,Cappelluti_prb12}  for bilayer graphene. Here the wavy lines denote the electromagnetic field, while solid and dashed  lines denote the  Green's functions for electrons and phonons, respectively. Labels at the vertices denote the insertion of either the velocity $v$ or the electron-phonon $V_{s\mu}$ operator. }	\label{Fig:diagrams}
\end{figure}
The enhancement of the ionic optical conductivity at the resonance condition $\Delta\simeq \omega_0$ in gapped disproportionated graphene is a remarkable result, because it implies a strong IR phonon activity, which is absent in pristine graphene. A similar mechanism has been observed so far only in bilayer graphene. In particular, by using a double-gate device able to control independently the charge doping and the electrostatic asymmetry between the layers, it has been shown 
in Ref.\ [\onlinecite{Feng_naturenano09}] that the IR activity of the zone-central phonon in undoped bilayer graphene is strongly enhanced when the bandgap is tuned to the phonon energy. A somehow related mechanism has been observed instead in single-gate devices, 
where both the bandgap and doping are simultaneously changed, leading to a sizable increase of the phonon strength with charge doping along with a marked evolution of the Fano profile\cite{Kuzmenko_prl09,Feng_naturenano09}. The explanation of these observations provided in Refs.~\citenum{Cappelluti_prb10,Cappelluti_prb12} within a diagrammatic field-theory scheme is completely analogous to the general scheme provided by Eqs..~(\ref{Eq:dynamic_charge}), (\ref{Eq:ionic_susceptibility}) and (\ref{Eq:oscillator_strength}). To make a closer connections between the two approaches, one should just take care of few technical differences. In the field-theory approaches one usually computes directly the complex optical conductivity $\bm\sigma(\omega)$ in linear response theory with respect to the gauge field $\bf A$, rather than the electrostatic potential. As a consequence, $\bm\sigma(\omega)$ is related to the electromagnetic tensor $K(\omega)$ relating the average total current to $\bf{A}$ . When one is interested only to the interband part, $\bm\sigma^{inter}(\omega)$ is directly given by the current-current correlation function, i.e.
$K_{\alpha\beta}(\omega)=e^2\langle v_\alpha v_\beta\rangle$, where $v_\alpha$ is the velocity operator defined in Eq.\ (\ref{defvertices}). In this approach, $\sigma^{inter}$ is again the sum of an electronic and ionic contribution. The electronic one $\sigma^{el}$ is in the diagrammatic formalism a bare bubble, see Fig.\ \ref{Fig:diagrams}, accounting only for the interband electronic transitions. The ionic contribution $\sigma^{ion}$ appears as a vertex correction to the bare bubble, and it describes all the possible (virtual or real) intermediate processes where the particle-hole excitation induced by the electric field decays in an intermediate phonon mode. The resulting contribution to the optical conductivity can be written as 
\begin{equation}
\sigma^{ion}_{\alpha\beta}(\omega)=-\sum_{s,\mu} \frac{K^s_{\alpha\mu}K^s_{\mu\beta}D_s(\omega)}
{ i\hbar \omega \Omega} 
\label{Eq:field}
\end{equation}
where $D_s(\omega)=2\hbar \omega_s/((\hbar \omega+i\gamma_s/2)-(\hbar \omega_s)^2)$ is the phonon propagator and the coupling between the electronic current and the phonon mode is encoded in the correlation functions $K_{\alpha\mu}^s=e\langle v_\alpha V_{s,\mu}\rangle$ and $K^s_{\beta\mu}=e\langle V^\dagger_{s,\mu}v_\beta\rangle$. Here $V_{s\mu}$ denotes the electron-phonon operator, which is given e.g. in the case of the $G$ mode of graphene by $\partial H_{\bf K}/{\partial u_\mu^A}$ defined in Eq.\ (\ref{defvertices}) above. By computing the contribution (\ref{Eq:field}) one easily recovers the expression (\ref{Eq:ionic_susceptibility}), with $f_{s\alpha}=(2\omega_s)^{-1/2}\sum_\mu K_{\alpha\mu}^s(\omega)/\omega$. In other words, one sees that the dynamical effective charge is represented in this formalism by the mixed current-phonon bubble $K^s_{\alpha\mu}$. Its structure is then strictly similar to Eq.\ (\ref{pijk}) above, i.e. it is a Lindhart-like response functions weighting with a symmetry-dependent factor  the  interband transitions $\epsilon_{\mathbf{k}i}-\epsilon_{\mathbf{k}j}$ coupled to the phonon. Since again the presence of the phonon propagator $D_s(\omega)$ in Eq.\ (\ref{Eq:field}) selects the value of the mixed bubble at $\omega=\omega_s$, in full analogy with the dynamical effective charge $Z(\omega_s)$ the $K_{\alpha\mu}(\omega_s)$ is real when the bandgap is larger than the phonon energy $\omega_s$, so that only virtual electronic excitations decay into the phonon, while it is imaginary when the $\Delta<\omega_s$ and real electron-hole pairs decay into the phonon.

\section{Conclusion}

In this paper we studied the polar responses of graphene when an external perturbation reduces its symmetry and induces the opening of an electronic bandgap. By computing ab-initio within a DFT approach the static response of gapped graphene we found two remarkable and somehow unexpected results. First, the piezoelectric coefficient and the Born effective charges are remarkably sizeable, independently on the symmetry-breaking mechanism, as soon as a finite bandgap opens. In the case of the effective charges they are even independent on $\Delta$ for $\Delta \lesssim 0.5$ eV, that is the experimentally relevant range. Second, the computed polar response is numerically giant and comparable to that of a fully polar material as h-BN.  We showed that the universal behavior of the piezoelectric and IR response in gapped graphene can be understood within the Dirac-like model description for the electronic degrees of freedom. Indeed, within this framework the coupling of electrons to the strain/lattice distortion can be always encoded as a coupling to an effective gauge field, so that both the piezoelectric coefficient and Born effective charge are simply proportional to the valley Chern number of Dirac electrons. The constant of proportionality, which can be readily estimated ab initio, is similar in graphene and h-BN, explaining the giant polar response of graphene and its fundamental independence on the gap for the experimentally accessible range of gap values. For larger gap values the analytical approximation breaks down and in particular the piezoelectric coefficient displays a marked suppression. As a result, the piezoelectric coefficient (per-layer) of gapped monolayer and bilayer graphene is three times larger than the one of h-BN, which is a large-gap polar material.

The ab-initio calculations have been limited, as usual, to the static response. On the other hand, when the band-gap becomes smaller than the relevant phononic energy scale the adiabatic approximation is expected to fail. While for the piezoelectric coefficient the phonons involved are acoustic ones, suppressing considerably the lower limit for the static approximation, in the case of the Born effective charge such lower limit is set already at $\omega_s\simeq 0.2$ eV, which is the energy of the optical $G$ mode. In order to describe the polar response in this low-gap regime we computed analytically the dynamical extension of the effective charge within the Dirac-like model, and we used it to explicitly compare the ionic conductivity to the electronic one for gapped graphene. The frequency-dependent calculation confirms that when $\Delta$ goes to zero the effective charge vanishes, in agreement with the symmetry-based expectation for ungapped graphene. However, along with this results two other remarkable phenomena are observed. First, the strength itself of the IR absorption displays a marked enhancement around the resonance condition $\Delta\simeq \omega_s$, leading to a huge phononic absorption. Second, as soon as the bandgap is smaller than $\omega_s$ a continuum of particle-hole excitations becomes available at the phonon energy, and the effective charge becomes a {\em complex} number. The interference between the real and imaginary part of the effective charges translates into the well-known Fano asymmetry of the phonon peak. These two findings, that were already discussed  in Refs. \onlinecite{Cappelluti_prb10,Cappelluti_prb12} within the context of a field-theory derivation of the optical properties of bilayer graphene, establish a direct link between the strength and Fano profile of the phononic absorption and the microscopic electronic mechanism responsible for the bandgap opening. 
All our findings  suggest that gaining control over the symmetry-breaking  mechanisms responsible for the band-gap opening will also offer a preferential tool to tune enormously its polar response.  In particular bilayer graphene, where the bandgap can be easily controlled in field-effect devices, represents the perfect playground to test the emergence of a giant polar response. Indeed,  the large piezoelectric response may have a potentially huge impact on  transport properties, due to the contribution of piezoelectric scattering to the electron mobility\cite{Price_ap80,Basu_jpc81}.  In addition, the possibility to implement  ab-initio the computation of the dynamical effective charge stands as a promising predictive tool able to  guide future experimental work on ad-hoc engineering of polar materials. 

\acknowledgments
We acknowledge financial support by the European Graphene Flagship Core 2 grant number 785219  and by the Italian MAECI under the Italian-India collaborative project  SUPERTOP-PGR04879. We also acknowledge the CINECA award under the ISCRA initiative (Grant HP10B3EDF2) for the availability of high performance computing resources.

\appendix

\section{Computational details} \label{App:DFT}
All calculations were performed using Quantum Espresso~\cite{0953-8984-21-39-395502} within the local-density approximation  (LDA)\cite{PhysRevLett.45.566}, adopting norm-conserving Trouiller-Martins-type pseudo-potentials \cite{PhysRevB.43.1993,*FUCHS199967} and a plane-wave expansion up to a 120 Ry cutoff.
The convergence threshold was set to $10^{-14}$ for self-consistent and phonon calculations and to $10^{-6}$ for Berry phase calculations.
An interlayer spacing of 14~\AA\ was adopted for monolayer graphene in order to avoid any correlation between layers. For gated bilayer graphene, instead, an interlayer distance of 3.35~\AA\ and a spacing between bilayers of 11.65~\AA\ were used. The bilayer graphene was placed in the middle of the cell, 
and the saw-like potential introduced to simulate the applied electric field increased in the region from 0.5~\AA\ to 13.5~\AA, with slope varying between 0 and 3.1~V/nm; 
in order to minimize edge effects, the change of the slope was located in the empty region.
Piezoelectric coefficients were calculated using an orthorhombic cell with 4 atoms per cell (8 in bilayer graphene), in order to align the direction of polarization along one of the lattice vectors, and a $2\%$ strain in the armchair direction ($y$). The orthorhombic structure was relaxed after the strain  deformation in order to minimize the forces on atoms. Since we were interested in the in-plane response, the atomic out-of-plane coordinates were kept fixed during relaxation in bilayer graphene. 
Brillouin-zone sampling was performed on $n\times n\times1$ Monkhorst-Pack~\cite{PhysRevB.13.5188} mesh with $n$ chosen on a case-by-case basis to ensure the convergence of the calculations. For disproportionated graphene, 
we used denser and denser $k$-point mesh as the 
bandgap gets smaller: $n=110$ for $\delta=0.1$ and 0.2, $n=80$ for $\delta$ between $0.3$ and $0.6$, $n=50$ for $\delta$ between $0.7$ and 1. For Distorted graphene, we used an energy cutoff of 100~Ry and fixed $n=80$ for all considered distortions but the smallest one ($\xi=\xi_0$), for which we used $n=100$. Eventually, in order to reach convergence in gated bilayer graphene, characterized by small gaps in the range 0-0.4~$eV$, it was necessary to use very dense grids for Brillouin zone sampling, up to $n=400$ for the smallest gap considered.

\section{Berry-phase formulation}\label{App:berry}

The electronic polarization, from which the electronic contribution to both the piezoelectric and Born effective charge tensors is derived, can be expressed for a two-dimensional system as\cite{PhysRevB.48.4442,RevModPhys.66.899}:
\begin{eqnarray}\label{def:Pberry}
\bm P(\pmb{\mathcal{A}}) &=&-e\sum_{s}\sum_v\int_{BZ}\frac{d{\bm k}}{(2\pi)^2}{\bm a_{ s}^{(v)}(\bm k, \pmb{\mathcal{ A}})},
\end{eqnarray}
where $s=\pm 1$ accounts for spin degrees of freedom, $v$ labels the occupied valence bands, $\pmb{\mathcal{ A}}$ is a gauge field describing the strain or deformation perturbation and the valence-band Berry connection is given by:
\begin{eqnarray}\label{eq:BC}
{\bm a_{s}^{(v)}(\bm k, \pmb{\mathcal{ A}})} &=& i\bra{u^{(v)}_{s}(\bm k,\pmb{\mathcal{ A}})}{\bm \nabla_k}\ket{u^{(v)}_{s}(\bm k,\pmb{\mathcal{ A}})},
\end{eqnarray}
$\ket{u^{(v)}_{s}(\bm k,\pmb{\mathcal{ A}})}$ being the eigenvector of the perturbed system. 

The polarization induced by the gauge perturbation $\pmb{\mathcal{ A}}$ in a 2D hexagonal lattice with $D_{3h}$ point-group symmetries can be generally written as
\begin{equation}
\bm P = \alpha \, \pmb{\mathcal{ A}}\times\hat{\bm z}
\end{equation}
which is equivalent, within linear response theory, with Eq. (\ref{Eq:definitions}) once the following definitions are provided:
\begin{subequations}
\begin{align}
\pmb{\mathcal{A}} = (\epsilon_{11}-\epsilon_{22})\,\hat{\bm x}\,-\,(\epsilon_{12}+\epsilon_{21})\,\hat{\bm y} && \alpha = e_{222} \\
\pmb{\mathcal{A}} = \hat{\bm z}\times\bm u\equiv -u_y\,\hat{\bm x}\,+ u_x\,\hat{\bm y}\quad\quad &&\quad \alpha = \frac{\vert e \vert}{\Omega}\, Z_\parallel \label{eq:bp_definitions}
\end{align}
\end{subequations}
With these definitions, and enforcing the symmetry requirements of the $D_{3h}$ point group, the linear-response coefficient can therefore be evaluated as\cite{Rostami_njp2d2018}:
\begin{eqnarray}\label{eq:BP_response}
\alpha &=&  -\frac{\partial P_y}{\partial \mathcal{A}_x} \equiv \frac{\partial P_x}{\partial \mathcal{A}}_y \nonumber\\
 &=& \frac{e}{2}N_{s}\int_{BZ}\frac{d{\bm k}}{(2\pi)^2} \left\{\frac{\partial a^{(v)}_{y}}{\partial \mathcal{A}_x} - \frac{\partial a^{(v)}_{x}}{\partial\mathcal{A}_y} \right\}_{\pmb{\mathcal{A}}\to 0},\nonumber\\
\end{eqnarray}
where $N_s=2$ is the spin degeneracy and
\begin{eqnarray}\label{eq:BCderivative}
\frac{\partial a^{(v)}_{\alpha}}{\partial \mathcal{A}_\beta} &=& - 2 Im \left\langle \frac{\partial }{\partial \mathcal{A}_\beta} u^{(v)}(\bm k,\pmb{\mathcal{ A}})\right.\left|\frac{\partial }{\partial k_\alpha} u^{(v)}(\bm k,\pmb{\mathcal{ A}})\right\rangle\nonumber\\
\end{eqnarray}
In order to evaluate Eq. (\ref{eq:BCderivative}), we introduce in analogy with Ref. \onlinecite{Rostami_njp2d2018} a fictitious velocity $\tilde{\bm v}$ defined as
\begin{equation}\label{eq:fictitous.vel}
\tilde{\bm v} = b^n \frac{\partial\mathcal{H}(\bm k,\pmb{\mathcal{A}})}{\partial \pmb{\mathcal{A}}}
\end{equation}
where the bond length $b$ is introduced to keep the correct units, being $n=1$ or $n=2$ for a strain or an atomic displacement perturbation, respectively. Using this fictitious velocity, both models for electron-strain and electron-lattice interactions can be recast as:
\begin{equation}
\mathcal{H} (\bm k,\pmb{\mathcal{A}}) = \mathcal{H}(\bm k) + \frac{1}{b^n}\sum_\alpha \mathcal{A}_\alpha \tilde{v}_\alpha(\bm k) 
\end{equation}
and the Bloch valence eigenvector of the perturbed two-bands system described by Eqs.\ (\ref{Eq:Hamiltonian-es}) and (\ref{Eq:Hamiltonian-el}) can be expressed, in first-order perturbation theory, as:
\begin{eqnarray}
\vert u^{\pi}_{\bm k,\pmb{\mathcal{A}}} \rangle &=& \ket{ u^{\pi}_{\bm k} } + \frac{1}{b^n} \frac{\langle u_{\bm k}^{\pi^\ast}\vert \tilde{\bm v}\cdot \pmb{\mathcal{A}} \vert  u^{\pi}_{\bm k} \rangle}{E^{\pi}_{\bm k}-E^{\pi^\ast}_{\bm k}}\vert  u^{\pi^\ast}_{\bm k} \rangle. \label{eq:pert_wf}
\end{eqnarray}
where eigenstates and eigenvalues of the unperturbed system are defined in Eqs.\ (\ref{Eq:eigenthings}).
Using Eq.\ (\ref{eq:pert_wf}) together with the following identity\cite{Xiao_berryreview}:
\begin{equation}
\left\langle u^{\pi^\ast}_{\bm k}\right\vert\left. \frac{\partial u^{\pi}_{\bm k}}{\partial k_\alpha}\right\rangle =
\frac{\langle u^{\pi^\ast}_{\bm k}\vert v_\alpha\vert u^{\pi}_{\bm k}\rangle}{E^{\pi}_{\bm k}-E^{\pi^\ast}_{\bm k}},
\end{equation}
where $\bm v$ is the velocity operator $ v_\alpha =\partial H_s/\partial k_\alpha$, we can express Eq.\ (\ref{eq:BCderivative}) as:
\begin{eqnarray}\label{eq:BC_perttheory}
\frac{\partial a^{\pi}_{\alpha}}{\partial \mathcal{A}_\beta} &=& - \frac{2}{b^n} Im \,\frac{\bra{u_{\bm k}^\pi} \tilde{v}_\beta\ket{u_{\bm k}^{\pi^\ast}}\bra{u_{\bm k}^{\pi^\ast}} v_\alpha\ket{u_{\bm k}^\pi}}{(E_{\bm k}^\pi-E_{\bm k}^{\pi^\ast})^2}.
\end{eqnarray}
The integrand of the linear-response coefficient in  Eq.\ (\ref{eq:BP_response}) then reads:
\begin{eqnarray}
\left\{\frac{\partial a^{(v)}_{y}}{\partial \mathcal{A}_x} - \frac{\partial a^{(v)}_{x}}{\partial\mathcal{A}_y} \right\} &=& -\frac{2}{b^n} Im \left\{\frac{\bra{u_{\bm k}^\pi} \tilde{v}_x\ket{u_{\bm k}^{\pi^\ast}}\bra{u_{\bm k}^{\pi^\ast}} v_y\ket{u_{\bm k}^\pi} }{(E_{\bm k}^\pi-E_{\bm k}^{\pi^\ast})^2} \right. \nonumber\\
&& - \,\left. \frac{\bra{u_{\bm k}^\pi} \tilde{v}_y\ket{u_{\bm k}^{\pi^\ast}}\bra{u_{\bm k}^{\pi^\ast}} v_x\ket{u_{\bm k}^\pi} }{(E_{\bm k}^\pi-E_{\bm k}^{\pi^\ast})^2}\right\}\nonumber\\
&=&\tilde{\Omega}^{\pi}(\bm k) + c.c.\equiv \tilde{\Omega}^{\pi}(\bm k)-\tilde{\Omega}^{\pi^\ast}(\bm k),\nonumber\\
\end{eqnarray}
where we introduced the generalized Berry curvatures for the valence and conduction  bands, respectively:
\begin{eqnarray}\label{Eq:genBerryCurv}
\tilde{\Omega}^\pi(\bm k) &=& \frac{i}{b^n}\,\frac{\bra{u_{\bm k}^\pi} \tilde{v}_x\ket{u_{\bm k}^{\pi^\ast}}\bra{u_{\bm k}^{\pi^\ast}} v_y\ket{u_{\bm k}^\pi} - (x\leftrightarrow y) }{(E_{\bm k}^\pi-E_{\bm k}^{\pi^\ast})^2} \nonumber \\
\tilde{\Omega}^{\pi^\ast}(\bm k) &=& \frac{i}{b^n}\,\frac{\bra{u_{\bm k}^{\pi^\ast}} \tilde{v}_x\ket{u_{\bm k}^{\pi}}\bra{u_{\bm k}^{\pi}} v_y\ket{u_{\bm k}^{\pi^\ast}} - (x\leftrightarrow y) }{(E_{\bm k}^\pi-E_{\bm k}^{\pi^\ast})^2} \nonumber \\
\end{eqnarray}
Notice that the contribution of the complex conjugate of $\tilde{\Omega}^\pi(\bm k)$ -- equal to the conduction-band generalized Berry curvature with opposite sign --  has been overlooked in Ref. \onlinecite{Rostami_njp2d2018}.

The Hamiltonian Eq.\ (\ref{Eq:Hamiltonian}), as well as the Hamiltonians Eqs.\ (\ref{Eq:Hamiltonian-es}) and (\ref{Eq:Hamiltonian-el}), describes the low-energy electronic bands at a given valley $K$; the corresponding Hamiltonian at valley $K'$ can be easily derived by imposing time-reversal symmetry. By introducing a valley index $\tau=\pm 1$, the Hamiltonians, including electron-strain and electron-lattice interaction, can be written in a compact form as $\mathcal{H}_\tau={\bm h}_\tau\cdot\bm\sigma$, where 
\begin{eqnarray}
{{\bm h}}_\tau^{e-s} &=& \hbar v_F \left(\tau k_x+\frac{\beta^{e-s}}{2b} \mathcal{A}^{e-s}_x,\,  k_y+\tau \frac{\beta^{e-s}}{2b} \mathcal{A}^{e-s}_y, \frac{\Delta}{2\hbar v_F}\right) \nonumber\\
\\
{{\bm h}}_\tau^{e-l} &=& \hbar v_F\left(\tau k_x-\frac{\beta^{e-l}}{b^2} u_y,\,  k_y+\tau \frac{\beta^{e-l}}{b^2} u_x, \frac{\Delta}{2 \hbar v_F}\right).\nonumber\\
\end{eqnarray}
Using these Hamiltonians, one immediately gets
\begin{subequations}
\begin{align}
\label{eq:fictitious.vstrain}
\tilde{\bm v}^{e-s} \,=\, \tau\, \frac{\beta^{e-s}}{2}\, \bm v &&\\
\label{eq:fictitious.vdispl}
\tilde{\bm v}^{e-l} \,=\,\tau \, \beta^{e-l}\,\bm v && 
\end{align}
\end{subequations}
Therefore, the generalized Berry curvatures Eqs.\ (\ref{Eq:genBerryCurv}) turn out to be proportional to the usual Berry curvatures $\Omega^\pi$, $\Omega^{\pi^\ast}$, which are strongly peaked around the valleys $K$, $K'$. Since in the considered two-bands models the relation $\Omega_\tau^\pi=-\Omega_\tau^{\pi^\ast}\equiv\Omega_\tau$ holds, we find $\Omega_\tau^\pi-\Omega_\tau^{\pi^\ast}=2 \Omega_\tau$, which explains the origin of the factor 2 missing in Eq.\ (8) of Ref. \onlinecite{Rostami_njp2d2018}. Plugging Eqs.\ (\ref{eq:fictitious.vstrain}),(\ref{eq:fictitious.vdispl}) in Eq.\ (\ref{Eq:genBerryCurv}) and decomposing the integral Eq.\ (\ref{eq:BP_response}) into its valley contributions, one recovers:
\begin{eqnarray}
e_{222} &=& \beta^{e-s}\,\frac{e}{4\pi b}\sum_\tau \tau \frac{1}{2\pi}\int d{\bm k}\, 2\,\Omega_\tau(\bm k) \nonumber\\
&\equiv& \beta^{e-s}\,\frac{e}{2\pi b}\sum_\tau \tau C_\tau \nonumber\\
\\
Z_\parallel &=& \beta^{e-l}\, \frac{\Omega}{4\pi b^2} \sum_\tau \tau \frac{1}{2\pi}\int d{\bm k}\, 2\,\Omega_\tau(\bm k) \nonumber\\ 
&\equiv&
\beta^{e-l}\,\frac{3 \sqrt{3}}{2\pi}\sum_\tau \tau C_\tau \nonumber\\
\end{eqnarray}
where the integral $\int_\tau$ is restricted around a given valley, with the valley-resolved Chern number being $C_\tau=\tau\sgn{\Delta}/2$.

\section{Electronic optical conductivity}	\label{App:conductivity}
The interband term of the electronic conductivity is calculated from DFPT as
\begin{equation}
\sigma_{xx}^{el}(\omega)=\sigma_0 \int_m^\infty \frac{dE}{2\pi} \left(\frac{1}{E}+\frac{\Delta^2}{4E^3}\right) \sum_{\zeta=\pm1}\frac{-2i\hbar\omega}{2E+\zeta\left(\hbar\omega+i\eta\right)}
\end{equation}
where $\sigma_0=\pi e^2/2h$ is the universal conductivity of clean graphene.
For infinite electronic lifetime ($\eta\rightarrow0$), the real part of $\sigma_{xx}^{el}(\omega)$ is thus given by
\begin{equation}
 Re[\sigma^{el}_{xx}(\omega)]=\sigma_0\left(1+\frac{\Delta^2}{(\hbar\omega)^2}\right)\theta\left(\frac{\hbar\omega}{2}-m\right).
\end{equation}
If we keep instead $\eta$ finite, the electronic contribution to the real part of the optical conductivity can be written as
\begin{align}
\sigma_{xx}^{el}(\omega)&=\sigma_0\frac{1}{\pi}\frac{(\hbar\omega)^2}{(\hbar\omega)^2+\eta^2} \left[\tilde\theta(m,\omega)-\frac{\eta}{\hbar\omega}\tilde\xi(m,\omega)\right]+ \nonumber \\
&+\sigma_0\frac{1}{\pi}\frac{(\hbar\omega)^4\Delta^2}{[(\hbar\omega)^2+\eta^2]^3} \left[\tilde\theta(m,\omega) - 3 \tilde\xi(m,\omega)\right]+ \nonumber \\
&+\sigma_0\frac{1}{\pi}\frac{(\hbar\omega)^2\Delta^2\eta^2}{[(\hbar\omega)^2+\eta^2]^3} \left[-3\tilde\theta(m,\omega) + \tilde\xi(m,\omega)\right]+ \nonumber \\
&+\sigma_0\frac{1}{\pi}\frac{(\hbar\omega)^2\Delta^2\eta}{2m[(\hbar\omega)^2+\eta^2]^2},
\end{align}
where $\tilde\theta(m,\omega)$ and $\tilde\xi(m,\omega)$ have already been defined in Eqs.~(\ref{Eq:theta-tilde}) and (\ref{Eq:csi-tilde}).

\section{Phonon linewidth}	\label{App:self-energy}
The optical-phonon self-energy of graphene at zero temperature has already been calculated in Ref.~\citenum{110005716689}.
It can be easily generalized to gapped graphene as
\begin{equation}
 \Pi(\omega)=-\lambda \int_m^\infty \frac{dE}{2\pi} \left(E+\frac{\Delta^2}{4E}\right) \sum_{\zeta=\pm1}\frac{1}{2E+\zeta\left(\hbar\omega+i\eta\right)}	\label{Eq:self-enery}
\end{equation}
where $m=\max(\Delta/2, E_F)$ and $\lambda$ is given by
\begin{equation}
 \lambda=\frac{9\sqrt{3}\hbar\beta^2}{M a^2 \omega_0}=\frac{4\bar \gamma}{\omega_0}\approx2.6 \times 10^{-2},
\end{equation}\\
where $\bar \gamma=9\sqrt{3}\hbar\beta^2/4Ma^2$.\\
The phonon line-width $\gamma^{e-ph}$ can be estimated as the self-energy evaluated at the phonon frequency, i.e.
\begin{equation}
 \gamma^{e-ph}=-2\frac{Im[\Pi(\omega_0)]}{\hbar}.
\end{equation}
For infinite electronic lifetime ($\eta\rightarrow0$), we get
\begin{equation}
 \gamma^{e-ph}= \bar \gamma  \left(1+\frac{\Delta^2}{(\hbar\omega_0)^2}\right) \theta\left(\frac{\hbar\omega_0}{2}-m\right).
\end{equation}
If we keep instead $\eta$ finite, the electron-phonon line-width can be written as
\begin{align}
\gamma^{e-ph} &= \bar \gamma  \left[\tilde\theta(m,\omega_0)+\frac{\eta}{\hbar\omega_0}\tilde\xi(m,\omega_0)\right]+ \nonumber \\
&+\bar \gamma\frac{1}{\pi} \frac{\Delta^2}{(\hbar\omega_0)^2+\eta^2}\left[\tilde\theta(m,\omega_0)-\frac{\eta}{\hbar\omega_0}\tilde\xi(m,\omega_0)\right],
\end{align}
where $\tilde\theta(m,\omega)$ and $\tilde\xi(m,\omega)$ have already been defined in Eqs. ~(\ref{Eq:theta-tilde}) and (\ref{Eq:csi-tilde}).

\end{document}